\documentclass[11pt,a4paper]{article}

\usepackage{jheppub}
\usepackage{amsfonts}
\usepackage{amssymb}
\usepackage{comment}
\usepackage{epsfig}
\usepackage{graphicx}
\usepackage{amsmath}
\usepackage{tabu}
\usepackage[all]{hypcap}

\newcommand{\bea}{\begin{eqnarray}}
\newcommand{\eea}{\end{eqnarray}}
\newcommand{\bi}{\begin{itemize}}
\newcommand{\ei}{\end{itemize}}
\newcommand{\ben}{\begin{enumerate}}
\newcommand{\een}{\end{enumerate}}
\newcommand{\be}{\begin{equation}}
\newcommand{\ee}{\end{equation}}
\newcommand{\ba}{\begin{align}}
\newcommand{\ea}{\end{align}}
\newcommand{\comments}[1]{}
\def\nn{\nonumber}

\def\SM{{\scriptscriptstyle \rm SM}}

\def\dS{{\scriptscriptstyle \rm dS}}

\def\DM{{\scriptscriptstyle \rm DM}}

\newcommand\vo{{\mathcal{V}}}

\newcommand{\mc}{\mathcal}
\newcommand{\ov}{\overline}

\newcommand{\beqa}{\begin{eqnarray}}
\newcommand{\eeqa}{\end{eqnarray}}
\newcommand{\V}{{\cal{V}}}

\title{General Analysis of Dark Radiation in Sequestered String Models}

\author[1,2,3]{Michele Cicoli,}
\author[2,3]{Francesco Muia}

\affiliation[1]{ICTP, Strada Costiera 11, Trieste 34014, Italy}
\affiliation[2]{Dipartimento di Fisica e Astronomia, Universit\`a di Bologna, \\ via Irnerio 46, 40126 Bologna, Italy}
\affiliation[3]{INFN, Sezione di Bologna, via Irnerio 46, 40126 Bologna, Italy}

\emailAdd{mcicoli@ictp.it}
\emailAdd{muia@bo.infn.it}

\abstract{We perform a general analysis of axionic dark radiation produced from the decay of the lightest modulus in the sequestered LARGE Volume Scenario. We discuss several cases depending on the form of the K\"ahler metric for visible sector matter fields and the mechanism responsible for achieving a de Sitter vacuum. The leading decay channels which determine dark radiation predictions are to hidden sector axions, visible sector Higgses and SUSY scalars depending on their mass. We show that in most of the parameter space of split SUSY-like models squarks and sleptons are heavier than the lightest modulus. Hence dark radiation predictions previously obtained for MSSM-like cases hold more generally also for split SUSY-like cases since the decay channel to SUSY scalars is kinematically forbidden. However the inclusion of string loop corrections to the K\"ahler potential gives rise to a parameter space region where the decay channel to SUSY scalars opens up, leading to a significant reduction of dark radiation production. In this case, the simplest model with a shift-symmetric Higgs sector can suppress the excess of dark radiation $\Delta N_{\rm eff}$ to values as small as $0.14$, in perfect agreement with current experimental bounds. Depending on the exact mass of the SUSY scalars all values in the range $0.14 \lesssim \Delta N_{\rm eff} \lesssim 1.6$ are allowed. Interestingly dark radiation overproduction can be avoided also in the absence of a Giudice-Masiero coupling.}


\begin{document}

\maketitle

\section{Introduction}

According to the cosmological Standard Model (SM), neutrinos were in thermal equilibrium at early times and decoupled at temperatures of order $1$ MeV. This decoupling left behind a cosmic neutrino background which has been emitted much earlier than the analogous cosmic microwave background (CMB). Due to the weakness of the weak interactions, this cosmic neutrino background cannot be detected directly, and so goes under the name of `dark radiation'. Its contribution to the total energy density $\rho_{\rm tot}$ is parameterised in terms of the effective number of neutrino-like species $N_{\rm eff}$ as:
\be
\rho_{\rm tot} = \rho_{\gamma} \left( 1 + \frac{7}{8}\left( \frac{4}{11} \right)^{4/3} N_{\rm eff} \right).
\ee
The SM predictions for $N_{\rm eff}$ are $N_{\rm eff}=3$ during Big Bang Nucleosynthesis (BBN) and $N_{\rm eff}=3.046$ at CMB times since neutrinos get slightly reheated when electrons and positrons annihilate. Any departure from these values would be a clear signal of physics beyond the SM due to the presence of extra dark radiation controlled by the parameter $\Delta N_{\rm eff}\equiv N_{\rm eff} - N_{\rm eff, SM}$.

Given that $N_{\rm eff}$ is positively correlated with the present value of the Hubble constant $H_0$, the comparison between indirect estimates of $H_0$ from CMB experiments and direct astrophysical measurements of $H_0$ could signal the need for extra dark radiation. The Planck 2013 value of the Hubble constant is $H_0 = (67.3\pm 1.2)$ km s$^{-1}$ Mpc$^{-1}$ ($68\%$ CL) \cite{Ade:2013zuv} which is in tension at $2.5\sigma$ with the Hubble Space Telescope (HST) value $H_0 = (73.8\pm 2.4)$ km s$^{-1}$ Mpc$^{-1}$ ($68\%$ CL) \cite{Riess:2011yx}. Hence the Planck 2013 estimate of $N_{\rm eff}$ with this HST `$H_0$ prior' is $N_{\rm eff} = 3.62^{+0.50}_{-0.48}$ ($95\%$ CL) \cite{Ade:2013zuv} which is more than $2\sigma$ away from the SM value and gives $\Delta N_{\rm eff} \leq 1.07$ at $2\sigma$.

However the HST Cepheid data have been reanalysed by \cite{Efstathiou:2013via} who found the different value $H_0 = (70.6\pm 3.3)$ km s$^{-1}$ Mpc$^{-1}$ ($68\%$ CL) which is within $1\sigma$ of the Planck 2015 estimate $H_0 = (67.3\pm 1.0)$ km s$^{-1}$ Mpc$^{-1}$ ($68\%$ CL) \cite{Planck:2015xua}. Hence the Planck 2015 collaboration performed a new estimate of $N_{\rm eff}$ without using any `$H_0$ prior' and obtaining $N_{\rm eff} = 3.13 \pm 0.32$ ($68\%$ CL) \cite{Planck:2015xua} which is perfectly consistent with the SM value and gives $\Delta N_{\rm eff} \leq 0.72$ at around $2\sigma$. This result might seem to imply that extra dark radiation is ruled out but this naive interpretation can be misleading since larger $N_{\rm eff}$ corresponds to larger $H_0$ and there is still an unresolved controversy in the direct measurement of $H_0$. In fact the Planck 2015 paper \cite{Planck:2015xua} analyses also the case with the prior $\Delta N_{\rm eff}=0.39$ obtaining the result $H_0 = (70.6\pm 1.0)$ km s$^{-1}$ Mpc$^{-1}$ ($68\%$ CL) which is even in better agreement with the new HST estimate of $H_0$ performed in \cite{Efstathiou:2013via}. Thus we stress that trustable direct astrophysical measurements of $H_0$ are crucial in order to obtain reliable bounds on $N_{\rm eff}$. 

$N_{\rm eff}$ is also constrained by measurements of primordial light element abundances. The Planck 2015 estimate of $N_{\rm eff}$ based on the helium primordial abundance and combined with the measurements of \cite{Aver:2013wba} is $N_{\rm eff}= 3.11^{+0.59}_{-0.57}$ ($95\%$ CL) giving $\Delta N_{\rm eff} \leq 0.65$ at $2\sigma$ \cite{Planck:2015xua}. However measurements of light element abundances are difficult and often affected by systematic errors, and so also in this case there is still some controversy in the literature since \cite{Izotov:2014fga} reported a larger helium abundance that, in turn, leads to $N_{\rm eff}= 3.58 \pm 0.50$ ($99\%$ CL) which is $3\sigma$ away from the SM value and gives $\Delta N_{\rm eff} \leq 1.03$ at $3\sigma$. Due to all these experimental considerations, in the rest of this paper we shall consider $\Delta N_{\rm eff}\lesssim 1$ as a reference upper bound for the presence of extra dark radiation.

Extra neutrino-like species can be produced in any beyond the SM theory which features hidden sectors with new relativistic degrees of freedom (\textit{dof}). In particular, extra dark radiation is naturally generated when reheating is driven by the decay of a gauge singlet since in this case there is no a priori good reason to suppress the branching ratio into hidden sector light particles \cite{DR1, DR2, AxionProbl}. 

This situation is reproduced in string models of the early universe due to the presence of gravitationally coupled moduli which get displaced from their minimum during inflation, start oscillating when the Hubble scale reaches their mass, quickly come to dominate the energy density of the universe since they redshift as matter and finally reheat the universe when they decay \cite{NTDM, NonThDMinSeqLVS}. In the presence of many moduli, the crucial one is the lightest since its decay dilutes any previous relic produced by the decay of heavier moduli.

Two important cosmological constraints have to be taken into account: (i) the lightest modulus has to decay before BBN in order to preserve the successful BBN predictions for the light element abundances \cite{CMP}; (ii) the modulus decay to gravitini should be suppressed in order to avoid problems of dark matter overproduction because of gravitini annihilation or modifications of BBN predictions \cite{gravProbl}. The first constraint sets a lower bound on the lightest modulus mass of order $m_{\rm mod} \gtrsim 30$ TeV, while a straightforward way to satisfy the second constraint is $m_{\rm mod}< 2 m_{3/2}$. 

However in general in string compactifications the moduli develop a mass because of supersymmetry (SUSY) breaking effects which make the gravitino massive via the super Higgs mechanism and generate also soft-terms of order $M_{\rm soft}$. Because of their common origin, one has therefore $m_{\rm mod}\sim m_{3/2}\sim M_{\rm soft}$. The cosmological lower bound $m_{\rm mod}\gtrsim 30$ TeV then pushes the soft-terms well above the TeV-scale ruining the solution of the hierarchy problem based on low-energy SUSY. An intriguing way-out is given by type IIB string compactifications where the visible sector is constructed via fractional D3-branes at singularities \cite{Aldazabal:2000sa, Conlon:2008wa, CYembedding}. In this case the blow-up modulus resolving the singularity is fixed at zero size in a supersymmetric manner, resulting in the absence of local SUSY breaking effects. SUSY is instead broken by bulk moduli far away from the visible sector singularity. Because of this geometric separation, the visible sector is said to be `sequestered' since the soft-terms can be suppressed with respect to the gravitino mass by $\epsilon = \frac{m_{3/2}}{M_P}\ll 1$ \cite{SeqLVS}. 

A concrete example of sequestered SUSY breaking is given by the type IIB LARGE Volume Scenario (LVS) with D3-branes at singularities, which is characterised by the following hierarchy of masses \cite{SeqLVS}:
\be
M_{1/2}\sim m_{3/2} \epsilon \ll m_{\rm mod} \sim m_{3/2} \sqrt{\epsilon} \ll m_{3/2}\,.
\label{masses}
\ee
This mass spectrum guarantees the absence of moduli decays to gravitini and allows for gaugino masses $M_{1/2}$ around the TeV-scale for $m_{\rm mod}\sim 10^7$ GeV and $m_{3/2}\sim 10^{10}$ GeV. On the other hand, SUSY scalar masses $m_0$ are more model dependent since their exact $\epsilon$-dependence is determined by the form of the K\"ahler metric for visible sector matter fields and the mechanism responsible for achieving a dS vacuum. The general analysis of \cite{SoftTermsSeqLVS} found two possible $\epsilon$-scalings for scalar masses: (i) $m_0\sim M_{1/2}$ corresponding to a typical MSSM-like scenario and (ii) $m_0 \sim m_{\rm mod} \gg M_{1/2}$ resulting in a split SUSY-like case with heavy squarks and sleptons.

Following the cosmological evolution of these scenarios, reheating takes place due to the decay of the volume modulus which produces, together with visible sector particles, also hidden sector \textit{dof} which could behave as extra dark radiation \cite{DR1, DR2}. Some hidden sector \textit{dof} are model dependent whereas others, like bulk closed string axions, are always present, and so give a non-zero contribution to $\Delta N_{\rm eff}$. In fact, as shown in \cite{DMDRcorr}, the production of axionic dark radiation is unavoidable in any string model where reheating is driven by the moduli decay and some of the moduli are stabilised by perturbative effects which keep the corresponding axions light. Note that light closed string axions can be removed from the low-energy spectrum via the St\"uckelberg mechanism only for cycles collapsed to zero size since in the case of cycles in the geometric regime the combination of axions eaten up by an anomalous $U(1)$ is mostly given by open string axions \cite{DMDRcorr}. 

R-parity odd visible sector particles produced from the lightest modulus decay subsequently decay to the lightest SUSY particle, which is one of the main dark matter candidates. Due to their common origin, axionic dark radiation and neutralino dark matter have an interesting correlation \cite{DMDRcorr}. In fact, by combining present bounds on $N_{\rm eff}$ with lower bounds on the reheating temperature $T_{\rm rh}$ as a function of the dark matter mass $m_\DM$ from recent Fermi data, one can set interesting constraints on the $(N_{\rm eff}, m_\DM)$-plane. \cite{DMDRcorr} found that standard thermal dark matter is allowed only if $\Delta N_{\rm eff}\to 0$ while the vast majority of the allowed parameter space requires non-thermal scenarios with Higgsino-like dark matter, in agreement with the results of \cite{Aparicio:2015sda} for the MSSM-like case.

Dark radiation production for the MSSM-like case has been studied in \cite{DR1, DR2} which showed that the leading decay channels of the volume modulus are to visible sector Higgses via the Giudice-Masiero (GM) term and to ultra-light bulk closed string axions. The simplest model with two Higgs doublets and a shift-symmetric Higgs sector yields $1.53 \lesssim \Delta N_{\rm eff}\lesssim 1.60$, where the window has been obtained by varying the reheating temperature between $500$ MeV and $5$ GeV, which are typical values for gravitationally coupled scalars with masses in the range $m_{\rm mod}\simeq (1 \div 5) \cdot 10^7$ GeV. These values of $\Delta N_{\rm eff}$ lead to dark radiation overproduction since they are in tension with current observational bounds.\footnote{Radiative corrections to the modulus coupling to Higgs fields do not give rise to a significant change in the final prediction for $\Delta N_{\rm eff}$ \cite{DRloop}.} Possible way-outs to reduce $\Delta N_{\rm eff}$ involve models with either a larger GM coupling or more than two Higgs doublets.

Due to this tension with dark radiation overproduction, different models have been studied in the literature. \cite{Angus:2014bia} showed how sequestered LVS models where the Calabi-Yau (CY) volume is controlled by more than one divisor are ruled out since they predict huge values of extra dark radiation of order $\Delta N_{\rm eff}\sim 10^4$. On the other hand, \cite{Hebecker:2014gka} focused on non-sequestered LVS models where the visible sector is realised via D7-branes wrapping the large cycle controlling the CY volume.\footnote{Another option involves flavour D7-branes wrapped around the volume divisor and intersecting the visible sector D3-branes localised at a singularity \cite{Hebecker:2014gka}.} In this way, the decay rate of the lightest modulus to visible sector gauge bosons becomes comparable to the decay to bulk axions, and so the prediction for $\Delta N_{\rm eff}$ can become smaller. In fact, the simplest model with a shift-symmetric Higgs sector yields $\Delta N_{\rm eff}\simeq 0.5$ \cite{Hebecker:2014gka}. However this case necessarily requires high-scale SUSY since without sequestering $M_{\rm soft}\sim m_{3/2}$ (up to loop factors), and so from~\eqref{masses} we see that the cosmological bound $m_{\rm mod}\sim m_{3/2} \sqrt{\epsilon}\gtrsim 30$ TeV implies $M_{\rm soft} \sim m_{3/2} \gtrsim \left(30\,{\rm TeV}\right)^{2/3} M_P^{1/3}\sim 10^9$ GeV. Moreover in this case the visible sector gauge coupling is set by the CY volume $\vo$, $\alpha_\SM^{-1} \sim \vo^{2/3} \sim 25$, and so it is hard to achieve large values of $\vo$ without introducing a severe fine-tuning of some underlying parameters. A possible way-out could be to consider anisotropic compactifications where the CY volume is controlled by a large divisor and a small cycle which supports the visible sector \cite{Cicoli:2011yy, Anis}.

In this paper we take instead a different point of view and keep focusing on sequestered models as in \cite{DR1, DR2} since they are particularly promising for phenomenological applications: they are compatible with TeV-scale SUSY and gauge coupling unification without suffering from any cosmological moduli and gravitino problem, they can be embedded in globally consistent CY compactifications \cite{CYembedding} and allow for successful inflationary models \cite{KMI} and neutralino non-thermal dark matter phenomenology \cite{Aparicio:2015sda}. Following the general analysis of SUSY breaking and its mediation to the visible sector performed in \cite{SoftTermsSeqLVS} for sequestered type IIB LVS models with D3-branes at singularities, we focus on the split-SUSY case where squarks and sleptons acquire a mass of order the lightest modulus mass: $m_0 = c\, m_{\rm mod}$ with $c\sim \mc{O}(1)$. We compute the exact value of the coefficient $c$ for different split-SUSY cases depending on the form of the K\"ahler metric for visible sector matter fields and the mechanism responsible for achieving a dS vacuum. We find that the condition $c\leq 1/2$, which allows the new decay channel to SUSY scalars, can be satisfied only by including string loop corrections to the K\"ahler potential \cite{Berg:2007wt, extendednoscale}. However this relation holds only at the string scale $M_s \sim 10^{15}$ GeV whereas the decay of the lightest modulus takes place at an energy of order its mass $m_{\rm mod}\sim 10^7$ GeV. Hence we consider the Renormalisation Group (RG) running of the SUSY scalar masses from $M_s$ to $m_{\rm mod}$ and then compare their value to $m_{\rm mod}$ whose running is in practice negligible since moduli have only gravitational couplings. Given that also the RG running of SUSY scalar masses is a negligible effect in split SUSY-like models, we find that radiative corrections do not alter the parameter space region where the lightest modulus decay to SUSY scalars opens up. 

We then compute the new predictions for $\Delta N_{\rm eff}$ which gets considerably reduced with respect to the MSSM-like case considered in \cite{DR1, DR2} since the branching ratio to visible sector particles increases due to the new decay to squarks and sleptons and the new contribution to the decay to Higgses from their mass term. We find that the simplest model with a shift-symmetric Higgs sector can suppress $\Delta N_{\rm eff}$ to values as small as $0.14$ in perfect agreement with current experimental bounds. Depending on the exact value of $m_0$ all values in the range $0.14 \lesssim \Delta N_{\rm eff} \lesssim 1.6$ are allowed. Interestingly $\Delta N_{\rm eff}$ can be within the allowed experimental window also in the case of vanishing GM coupling $Z = 0$ since the main suppression of $\Delta N_{\rm eff}$ comes from the lightest modulus decay to squarks and sleptons. Given that a correct realisation of radiative Electro-Weak Symmetry Breaking (EWSB) in split SUSY-like models requires in general a large $\mu$-term of order $m_0$, the lightest modulus branching ratio into visible sector \textit{dof} is also slightly increased due to its decay to Higgsinos. However this new decay channel yields just a negligible correction to the final prediction for dark radiation production. 

This paper is organised as follows. In Sec.~\ref{LVSs} we review the main features of sequestered LVS models whereas Sec.~\ref{drs} contains the main results of this paper since it analyses the predictions for axionic dark radiation. We present our conclusions in Sec.~\ref{Concl}.

\section{Sequestered LARGE Volume Scenario}
\label{LVSs}

In this section we shall present a brief review of sequestered type IIB LVS models with D3-branes at singularities \cite{CYembedding, SeqLVS, SoftTermsSeqLVS}. After describing the general setup of the $N=1$ supergravity effective field theory, we outline the procedure followed to fix all closed string moduli in a dS vacuum which breaks SUSY spontaneously. We then list all the relevant D- and F-terms and the final results for the soft-terms generated by gravity mediation (anomaly mediation contributions can be shown to be negligible \cite{SeqLVS, SoftTermsSeqLVS}).

\subsection{Effective field theory setup}

The simplest LVS model which leads to a visible sector sequestered from SUSY breaking is based on a CY manifold whose volume takes the form:
\be
\vo = \left(T_b+\overline{T}_b\right)^{3/2} - \left(T_s+\overline{T}_s\right)^{3/2} - \left(T_\SM+\overline{T}_\SM\right)^{3/2}\,,
\ee
where $\tau_i~\equiv {\rm Re} (T_i) = {\rm Vol}(D_i)$ parameterises the volume of the divisor $D_i$ and $\psi_i~\equiv {\rm Im}(T_i) = \int_{D_i} C_4$. $D_b$ is the `big' divisor which controls the CY size, $D_s$ is a `small' blow-up modulus supporting non-perturbative effects and $D_\SM$ is the `Standard Model' cycle which collapses to a singularity where the visible sector D3-branes are localised. More precisely, explicit realisations in compact CY manifolds involve two identical shrinkable divisors $D_1$ and $D_2$ which are exchanged by the orientifold involution \cite{CYembedding}. This gives rise to an orientifold even cycle $D_+=D_1+D_2$ and an orientifold odd cycle $D_-=D_1-D_2$. $D_+$ is identified with $D_\SM$ whereas $D_-$ gives rise to an additional K\"ahler modulus $G= \int_{D_-} B_2 +i \int_{D_-} C_2$. Both $\tau_\SM$ and ${\rm Re}(G)$ develop a vanishing VEV due to D-term stabilisation while the corresponding axions $\psi_\SM$ and ${\rm Im}(G)$ are eaten up by two anomalous $U(1)$s \cite{Conlon:2008wa}. 

The resulting low-energy $N=1$ effective field theory is characterised by the following K\"ahler potential with leading order $\alpha'$ correction (without loss of generality we ignore from now on the orientifold odd modulus $G$):
\be
K = - 2 \ln\left(\vo + \frac{\hat\xi}{2}\right) + \frac{\tau_\SM^2}{\vo} - \ln\left(S+\overline{S}\right) + K_{\rm cs}(U) + K_{\rm matter}\,,
\label{generalk}
\ee
where $S$ is the axion-dilaton, $\hat\xi\equiv \xi {\rm Re}(S)^{3/2}$, $K_{\rm cs}(U)$ is the tree-level K\"ahler potential for the complex structure moduli $U$ and $K_{\rm matter}$ is the K\"ahler potential for visible matter fields $C$ which is taken to be:
\be
K_{\rm matter} = \tilde{K}\left[f_\alpha(U,S)\,\ov{C}^{\ov{\alpha}} C^\alpha + Z(U,S)\,\left(H_u H_d + {\rm h.c.}\right)\right]\,,
\label{kmatter}
\ee
where the matter metric is assumed to be flavour diagonal and \cite{SoftTermsSeqLVS}: 
\be
\tilde{K}=\frac{1}{\vo^{2/3}}\left(1-c_s\frac{\hat\xi}{\vo}\right),
\label{Ktilde}
\ee
and the bilinear Higgs mixing term is proportional to the GM coupling $Z$ which we allow to depend on $S$ and $U$-moduli. Note that additional contributions to~\eqref{generalk} can come from either an extra sector responsible for achieving a dS vacuum or from higher $\alpha'$ and $g_s$ corrections \cite{Berg:2007wt, extendednoscale}.

The superpotential takes instead the form:
\be
W = W_{\rm flux}(U,S) + A(U,S)\,e^{-a_s T_s} + W_{\rm matter}\,,
\label{sp}
\ee
where $W_{\rm flux}(U,S)$ is generated by three-form background fluxes, $A(U,S)$ and $a_s$ depend on the D-brane configuration which generates non-perturbative effects and the matter superpotential $W_{\rm matter}$ takes the form:
\be
W_{\rm matter} = \mu(U,S,T) H_u H_d + Y_{\alpha \beta \gamma}(U,S,T)  C^\alpha C^\beta C^\gamma + \dots\,,
\ee
where the $\mu$-term and the Yukawa couplings $Y_{\alpha \beta \gamma}$ can depend on the $T$-moduli only at non-perturbative level.

Finally, the expression for the gauge kinetic function of the visible sector localised at the singularity $\tau_\SM\to 0$ is:
\be
f_a = k_a S+ \lambda_a T_\SM\,\to\, k_a S\,,
\label{gkf}
\ee
where $k_a$ is a singularity-dependent coefficient.

\subsection{de Sitter moduli stabilisation}
\label{mss}

The scalar potential receives several contributions from various effects which are suppressed by different inverse powers of the overall volume $\vo$. By taking the large volume limit $\vo\gg 1$, the moduli can therefore be stabilised order by order in $1/\vo$. All closed string moduli are stabilised $\acute{a}$ la LVS by the following procedure:

\bi
\item At leading $1/\vo^2$ order, the dilaton and the $U$-moduli are stabilised by background fluxes while $\tau_\SM$ shrinks to zero via D-term stabilisation and $\psi_\SM$ is eaten up by an anomalous $U(1)$ \cite{Conlon:2008wa, CYembedding}.

\item At subleading $1/\vo^3$ order, $\tau_s$ and $\psi_s$ are stabilised by non-perturbative corrections to $W$ while $\tau_b$ is fixed at exponentially large values by the interplay between $\alpha'$ and non-perturbative effects. The scalar potential at $1/\vo^3$ order looks like:
\be
V = \frac{8}{3}(a_s A_s)^2 \sqrt{\tau_s} \,\frac{e^{- 2 a_s \tau_s}}{\vo}- 4 a_s A_s W_0 \tau_s \,\frac{e^{-a_s \tau_s}}{\vo^2 } + \frac{3\hat\xi W_0^2}{4\vo^3} + V_\dS\,,
\label{v0}
\ee
where we included a model-dependent positive contribution $V_\dS$ in order to obtain a dS solution. The minimum of this potential is at:
\be
e^{-a_s \tau_s} \simeq \frac{3 \sqrt{\tau_s} W_0}{4 a_s A_s \vo} \qquad\qquad \tau_s^{3/2} \simeq \frac{\hat{\xi}}{2}= \frac{\xi}{2}\frac{1}{g_s^{3/2}}\,,
\label{lvsmin}
\ee
with subleading $V_\dS$-dependent corrections. Following \cite{SoftTermsSeqLVS}, we shall consider two possible mechanisms to generate $V_\dS$:
\ben
\item dS$_1$ case: hidden sector matter fields $\phi_\dS$ living on D7-branes wrapping $D_b$ acquire non-zero VEVs because of D-term stabilisation. In turn, their F-term scalar potential gives rise to a positive contribution which scales as $V_\dS \sim W_0^2/\vo^{8/3}$ and can be used to obtain a dS vacuum \cite{CYembedding}. 
\item dS$_2$ case: a viable dS vacuum can arise also due to non-perturbative effects at the singularity obtained by shrinking to zero an additional divisor $D_\dS$ \cite{Cicoli:2012fh}. The corresponding K\"ahler modulus $T_\dS = \tau_\dS + i \psi_\dS$ is again fixed by D-terms which set $\tau_\dS\to 0$ ($\psi_\dS$ is eaten up by an anomalous $U(1)$). The new contribution to the superpotential $W_\dS = A_\dS e^{-a_\dS (S + k_\dS T_\dS)}$ yields a positive term which scales as $V_\dS \sim e^{-2 a_\dS {\rm Re}(S)}/\vo$ and can be used to develop a dS vacuum.
\een

\item At very suppressed $e^{-\vo^{2/3}}/\vo^{4/3}$ order, also $\psi_b$ develops a mass via $T_b$-dependent non-perturbative contributions to $W$. For $\vo\gtrsim 5\cdot 10^3$, the scale of this tiny contribution to $V$ is smaller than the present value of the cosmological constant, and so no additional `uplifting' term is needed.
\ei

\subsection{F- and D-terms}
\label{sts}

The typical feature of models with D3-branes at singularities is the fact that the SM modulus $T_\SM$ (together with the corresponding orientifold odd modulus $G$) does not break SUSY since its F-term is proportional to $\tau_\SM$ that shrinks to zero via D-term stabilisation:
\be
F^{T_\SM} = F^G = 0\,.
\ee
Therefore there is no local SUSY breaking effect and the visible sector is said to be \textit{sequestered} from SUSY breaking which takes place in the bulk via the following non-zero F-terms (we write down only the leading order expressions) \cite{SoftTermsSeqLVS}:
\be
\frac{F^{T_b}}{\tau_b} \simeq -2 m_{3/2}\,, \qquad \qquad \frac{F^{T_s}}{\tau_s} \simeq -\frac 32 \frac{m_{3/2}}{a_s\tau_s}\,, 
\ee
where the gravitino mass is $m_{3/2} = e^{K/2} |W| \sim M_P/\vo$. Because of subleading $S$ and $U$-dependent corrections to $V$ at $1/\vo^3$ order, also the dilaton and complex structure moduli develop non-zero F-terms of order \cite{SoftTermsSeqLVS}:
\be
\frac{F^S}{s}\sim \frac{m_{3/2}}{\vo}\left(\ln\vo\right)^{3/2}\qquad\text{and}\qquad F^U\sim F^S\,.
\ee
Moreover, the fields responsible for achieving a dS vacuum contribute to SUSY breaking via \cite{SoftTermsSeqLVS}:
\be
{\rm dS}_1 \, {\rm case}: \quad F^{\phi_\dS} \simeq \phi_\dS m_{3/2}  \qquad\qquad {\rm dS}_2 \,{\rm case}: \quad F^{T_\dS} \simeq m_{3/2}\,.
\ee
Further contributions to SUSY breaking come from non-zero D-terms. For non-tachyonic scalar masses, the visible sector D-term potential vanishes whereas for the two dS cases, the hidden sector D-term potential at the minimum scales as \cite{SoftTermsSeqLVS}:
\be
{\rm dS}_1 \, {\rm case}: \quad V_D \sim m_{3/2}^4 \vo^{2/3} \sim \vo^{-10/3}  \qquad\qquad {\rm dS}_2 \,{\rm case}: \quad V_D\sim m_{3/2}^4 \sim \vo^{-4}\,.
\ee

\subsection{Soft SUSY breaking terms}

Gravitational interactions mediate SUSY breaking to the visible sector generating soft-terms at the string scale. Let us summarise all the main results for the soft-terms.

\subsubsection{Gaugino masses}

Given the form of the gauge kinetic functions~\eqref{gkf}, the gaugino masses turn out to be universal and read:
\be
M_{1/2} = \frac{M_P}{2 \text{Re}(f)}\, F^i\partial_i f = \frac{F^S}{2 s} \sim \frac{m_{3/2}}{\vo}\left(\ln\vo\right)^{3/2}\,.
\label{gm}
\ee

\subsubsection{Scalar masses}
\label{Sec:m0}

Scalar masses depend on the K\"ahler matter metric since their supergravity expression reads:
\be
m_\alpha^2 = m_{3/2}^2 - F^i \ov{F}^{\ov{j}} \partial_i \partial_{\ov{j}} \ln \tilde{K}_\alpha\,.
\ee
It is therefore crucial to determine the exact moduli-dependence of $\tilde{K}_\alpha$ focusing in particular on the dependence on the volume $\vo$ since $F^{T_b}$ is the largest F-term. This dependence can be inferred relatively easily by requiring that the physical Yukawa couplings $\hat{Y}_{\alpha \beta \gamma}$ (we neglect here tiny non-perturbative $T$-dependent contributions to $Y_{\alpha \beta \gamma}$):
\be
\label{yuk}
\hat{Y}_{\alpha \beta \gamma} = e^{K/2}\,\frac{Y_{\alpha \beta \gamma}(U,S)}{\sqrt {\tilde{K}_\alpha \tilde{K}_\beta \tilde{K}_\gamma} }
= \left(\frac{e^K}{\tilde{K}^3}\right)^{1/2}\frac{Y_{\alpha \beta \gamma}(U,S)}{\sqrt {f_\alpha f_\beta f_\gamma} } \,,
\ee
should not depend on $\vo$ due to the locality of the SM construction. This translates into the condition:
\be
\tilde{K} = e^{K/3}\,,
\label{ul}
\ee
which can lead to two different limits \cite{SoftTermsSeqLVS}:
\bi
\item[1.] \textit{Local limit}: the relation~\eqref{ul} holds only at leading order in $1/\vo$;
\item[2.] \textit{Ultra-local limit}: the relation~\eqref{ul} holds exactly.
\ei
The expansion~\eqref{Ktilde} of $\tilde{K}$ guarantees that~\eqref{ul} is exact at leading order, as required in the local case. The ultra-local limit is then realised only for $c_s = 1/3$. Let us see how scalar masses get affected by the exact form of $\tilde{K}$:
\bi
\item \emph{Local limit}: in this case scalar masses are universal and do not depend on the sector responsible for getting a dS vacuum. Their leading order expression is generated by $F^{T_b}$ and looks like:
\be
m_0^2 = \frac{15}{2}\left(c_s-\frac{1}{3}\right)\, \frac{m_{3/2}^2 \tau_s^{3/2}}{\vo}\sim m_{3/2} M_{1/2}\,,
\label{lsm}
\ee
Note that non-tachyonic scalars require $c_s \geq 1/3$ and $M_{1/2}\ll m_0$ leading to a typical split-SUSY spectrum.

\item \emph{Ultra-local limit}: in this case scalar masses depend on the dS sector. Setting $c_s=1/3$ kills the leading contribution to~\eqref{lsm} from $F^{T_b}$. Subleading effects depend on $F^U$ and $F^S$ which are volume-suppressed with respect to $F^{T_b}$ since the dilaton and complex structure moduli are fixed supersymmetrically at leading order. Due to this cancellation, D-term contributions to scalar masses turn out to be the dominant effect for the dS$_1$ case leading to universal and non-tachyonic scalar masses of the form:
\be
{\rm dS}_1 \, {\rm case}: \qquad m_0^2 = \frac{9}{4 a_s\tau_s} \frac{m_{3/2}^2\tau_s^{3/2}}{\vo}\sim \frac{m_{3/2}}{\ln\vo}M_{1/2}\,.
\label{ul1}
\ee
This is again a split SUSY-like scenario. On the other hand, D-term contributions vanish in the dS$_2$ case where the leading effects generating scalar masses come from $F^U$ and $F^S$ that give:
\be
{\rm dS}_2 \, {\rm case}: \qquad \quad m_\alpha^2 =  c_\alpha(U,S) {\rm Re}(S)^2 M_{1/2}^2\,,
\label{ul2}
\ee
where the function $c_\alpha(U,S)$ involves derivatives of $f_\alpha(U,S)$ with respect to $U$ and $S$ \cite{SoftTermsSeqLVS}. In this case scalar masses are potentially non-universal and tachyonic depending on the exact functional dependence of $f_\alpha(U,S)$. This situation reproduces a standard MSSM-like scenario. Dark radiation production in this case has been studied in \cite{DR1, DR2}.
\ei

\subsubsection{$\mu$ and $B\mu$ terms}

The other soft-terms relevant for the computation of dark radiation production from moduli decays are the canonically normalised $\hat\mu$ and $B\hat{\mu}$ terms. These two terms receive contributions from both the K\"ahler potential and the superpotential. The more model-independent contribution from $K$ is induced by a non-zero GM coupling $Z$ in~\eqref{kmatter}. It turns out that in each dS case $\hat\mu$ is always proportional to $M_{1/2}$ whereas $B\hat\mu$ scales as $m_0^2$ \cite{SoftTermsSeqLVS}:
\be
\hat\mu = c_{\mu,K}(U,S) Z M_{1/2}\,, \qquad \qquad B \hat\mu = c_{B,K}(U,S) Z m_0^2\,,
\label{bmu}
\ee
where $c_{\mu,K}(U,S)$ and $c_{B,K}(U,S)$ are two tunable flux-dependent coefficients.

Additional contributions to $\hat{\mu}$ and $B\hat{\mu}$ can come from model-dependent non-perturbative effects which produce an effective $\mu$-term in the superpotential of the form \cite{SoftTermsSeqLVS}:
\be
W \supset e^{-a T} H_u H_d\,,
\ee
if the cycle associated to $\tau = \text{Re}(T)$ is in the geometric regime while:
\be
W \supset e^{-b(S + \kappa T)} H_u H_d\,,
\ee
if the cycle associated to $\tau = \text{Re}(T)$ is in the singular regime, i.e. if $\tau \rightarrow 0$. These non-perturbative effects generate effective $\hat{\mu}$- and $B\hat{\mu}$-terms in the low-energy action which can be parameterised as:
\be
\label{mubmu}
\hat{\mu} \simeq \frac{c_{\mu,W}(U,S)}{\V^{n + \frac{1}{3}}}\,, \qquad\qquad B\hat{\mu} \simeq \frac{c_{B,W}(U,S)}{\V^{n + \frac{4}{3}}}\,,
\ee
where $c_{\mu,W}(U,S)$ and $c_{B,W}(U,S)$ are flux-dependent tunable coefficients and $n$ is the instanton number. As we shall explain in Sec.~\ref{splitsusycase}, this model-dependent contribution to the $\mu$-term is crucial to reproduce a correct radiative EWSB for most of the parameter space of split SUSY-like models.

\section{Dark radiation in sequestered models}
\label{drs}

As already argued in the Introduction, the production of dark radiation is a generic feature of string models where some of the moduli are fixed by perturbative effects \cite{DMDRcorr}. In fact, if perturbative corrections fix the real part of the modulus $T=\tau + {\rm i}\psi$, the axion $\psi$ remains exactly massless at this level of approximation due to its shift symmetry, leading to $m_\tau \gg m_\psi$. Hence very light relativistic axions can be produced by the decay of $\tau$, giving rise to $\Delta N_{\rm eff} \neq 0$. 

\subsection{Dark radiation from moduli decays}

Following the cosmological evolution of the Universe, during inflation the canonically normalised modulus $\Phi$ gets a displacement from its late-time minimum of order $M_P$. After the end of inflation the value of the Hubble parameter $H$ decreases. When $H \sim m_\Phi$, $\Phi$ starts oscillating around its minimum and stores energy. During this stage $\Phi$ redshifts as matter, so that it quickly comes to dominate the energy density of the Universe. Afterwards reheating is caused by the decay of $\Phi$ which takes place when:
\be
3 H^2 \simeq \frac43 \,\Gamma_\Phi^2\,,
\label{HG}
\ee
where $\Gamma_\Phi$ is the total decay rate into visible and hidden \textit{dof}:
\be
\Gamma_\Phi = \Gamma_{\rm vis} + \Gamma_{\rm hid} = \left(c_{\rm vis} + c_{\rm vis}\right) \Gamma_0\,,\qquad\text{with}
\qquad \Gamma_0~\equiv \frac{1}{48\pi}\frac{m_\Phi^3}{M_P^2}\,.
\label{G}
\ee
The corresponding reheating temperature is given by:
\be 
T_{\rm rh} = \left(\frac{30\,\rho_{\rm vis}}{\pi^2 g_*(T_{\rm rh})}\right)^{\frac 14}\,,
\label{Trh}
\ee
where $\rho_{\rm vis} = \left(c_{\rm vis}/c_{\rm tot}\right) 3 H^2 M_P^2$ with $c_{\rm tot}=c_{\rm vis} + c_{\rm hid}$. 
Using~\eqref{HG} and~\eqref{G} $T_{\rm rh}$ can be rewritten as:
\be
T_{\rm rh} \simeq \frac{1}{\pi} \left(\frac{5 c_{\rm vis} c_{\rm tot}}{288 g_*(T_{\rm rh})}\right)^{1/4} m_\Phi\,\sqrt{\frac{m_\Phi}{M_P}}\,.
\label{rt}
\ee
This reheating temperature has to be larger than about $1$ MeV in order to preserve the successful BBN predictions. 

In the presence of a non-zero branching ratio for $\Phi$ decays into hidden sector \textit{dof}, i.e. for $c_{\rm hid}\neq 0$, extra axionic dark radiation gets produced, leading to \cite{DR1, DR2}:
\be
\Delta N_{\rm eff} = \frac{43}{7} \frac{c_{\rm hid}}{c_{\rm vis}} \left(\frac{g_*(T_{\rm dec})}{g_*(T_{\rm rh})}\right)^{1/3},
\label{dn}
\ee
where $T_{\rm dec} \simeq 1$ MeV is the temperature of the Universe at neutrino decoupling with $g_*(T_{\rm dec}) = 10.75$. 
The factor in brackets is due to the fact that axions are very weakly coupled (they are in practice only gravitationally coupled), and so they never reach thermal equilibrium. Therefore, given that the comoving entropy density $g_*(T) T^3 a^3$ is conserved, the thermal bath gets slightly reheated when some species drop out of thermal equilibrium. Note that the observational reference bound $\Delta N_{\rm eff}\lesssim 1$ implies:
\be
c_{\rm vis}\gtrsim 3\, c_{\rm hid}\qquad\text{for}\qquad T_{\rm rh}\gtrsim 0.2\,\text{GeV}\,,
\label{cbound}
\ee
where we have used the fact that $g_*(T_{\rm rh}) = 75.75$ in the window $0.2\,{\rm GeV} \lesssim T_{\rm rh} \lesssim 0.7\, {\rm GeV}$ while $g_*(T_{\rm rh}) = 86.25$ for $T_{\rm rh} \gtrsim 0.7\,\rm GeV$.

\subsection{Light relativistic axions in LVS models}

Let us summarise the main reasons why axionic dark radiation production is a typical feature of sequestered LVS models:
\bi
\item Reheating is driven by the last modulus to decay which is $\tau_b$ since the moduli mass spectrum takes the form (the axion $\psi_\SM$ is eaten up by an anomalous $U(1)$):
\be
m_{\tau_b}\sim m_{3/2} \sqrt{\epsilon} \ll m_{\tau_s} \sim m_{\psi_s} \sim m_S \sim m_U \sim m_{3/2} \ll m_{\tau_\SM} \sim \frac{m_{3/2}}{\sqrt{\epsilon}}\sim M_s\,,
\ee
where $\epsilon = m_{3/2}/M_P \sim W_0 / \vo \ll 1$. Given that gaugino masses scale as $M_{1/2}\sim m_{3/2} \epsilon$, TeV-scale SUSY fixes $m_{\tau_b}$ around $10^7$ GeV which in turn, using~\eqref{rt}, gives $T_{\rm rh}$ around $1$ GeV.\footnote{As in standard split SUSY models, we require TeV-scale gauginos for dark matter and gauge coupling unification. In MSSM-like models we focus on low-energy SUSY to address the hierarchy problem.} Note that $m_{\tau_b}\ll m_{3/2}$, and so sequestering addresses the gravitino problem since the decay of the volume modulus into gravitinos is kinematically forbidden.

\item Given that axions enjoy shift symmetries which are broken only by non-perturbative effects, the axionic partner $\psi_b$ of the volume mode $\tau_b$ is stabilised by non-perturbative contributions to the superpotential of the form $W \supset A_b\, e^{- a_b T_b} \sim e^{-\vo^{2/3}}\ll 1$. These tiny effects give rise to a vanishingly small mass $m_{\psi_b}^2 \sim e^{- \vo^{2/3}} \sim 0$. Hence these bulk closed string axions are in practice massless and can be produced from the decay of $\tau_b$ \cite{DR1, DR2}. 

\item Some closed string axions can be removed from the 4D spectrum via the St\"uckelberg mechanism in the process of anomaly cancellation. However, the combination of bulk axions eaten up by an anomalous $U(1)$ is mostly given by an open string mode, and so $\psi_b$ survives in the low-energy theory (the situation is opposite for axions at local singularities) \cite{DMDRcorr}. 
\ei

\subsection{Volume modulus decay channels}

The aim of this section is to compute the ratio $c_{\rm hid}/c_{\rm vis}$ which is needed to predict the effective number of extra neutrino-like species $\Delta N_{\rm eff}$ using (\ref{dn}).

\subsubsection{Decays into hidden sector fields}
\label{vmds}

Some hidden sector \textit{dof} are model dependent whereas others are generic features of LVS models. As pointed out above, bulk closed string axions are always a source of dark radiation. On top of them, there are local closed string axions which however tend to be eaten up by anomalous $U(1)$s (this is always the case for each del Pezzo singularity) and local open string axions (one of them could be the QCD axion \cite{CYembedding}) whose production from $\tau_b$ decay is negligibly small \cite{DR1}. Moreover the decay of $\tau_b$ into bulk closed string $U(1)$s is also a subdominant effect \cite{DR1}. Model dependent decay channels involve light \textit{dof} living on hidden D7-branes wrapping either $D_b$ or $D_s$ and hidden D3-branes at singularities which are geometrically separated from the one where the visible sector is localised. However, as explained in \cite{DR1}, the only decay channels which are not volume or loop suppressed are to light gauge bosons on the large cycle and to Higgses living on sequestered D3s different from the visible sector. Given that the presence of these states is non-generic and can be avoided by suitable hidden sector model building, we shall focus here just on $\tau_b$ decays into 
bulk closed string axions. 

The corresponding decay rate takes the form \cite{DR1,DR2}: 
\be
\Gamma_{\Phi \rightarrow a a} = \Gamma_0\qquad\Rightarrow\qquad c_{\rm hid}=1\,,
\label{ddr}
\ee
where $\Phi$ and $a$ are, respectively, the canonically normalised real and imaginary parts of the big modulus $T_b$.
This result can be derived from the tree-level K\"ahler potential:
\be
K \simeq - 3 \ln\left(\frac{T_b+\overline{T}_b}{2}\right)\,,
\ee
which gives a kinetic Lagrangian of the form:
\be
\mc{L}_{\rm kin} = \frac{3}{4 \tau_b^2} \left(\partial_\mu \tau_b \partial^\mu \tau_b + \partial_\mu \psi_b \partial^\mu \psi_b\right)\,.
\label{lag1}
\ee
After canonical normalisation of $\tau_b$ and $\psi_b$:
\be
\label{cn1}
\frac{\Phi}{M_P} = \sqrt{\frac{3}{2}} \ln \tau_b\,, \qquad \qquad \frac{a}{M_P} = \sqrt{\frac{3}{2}} \frac{\psi_b}{\langle \tau_b \rangle}\,,
\ee
and expanding $\Phi$ as $\Phi = \Phi_0 + \hat \Phi$, the kinetic Lagrangian~\eqref{lag1} can be rewritten as:
\be
\mc{L}_{\rm kin} = \frac 12 \partial_\mu \hat\Phi \partial^\mu \hat\Phi + \frac 12 \partial_\mu a \partial^\mu a 
- \sqrt{\frac 23 } \frac{\hat \Phi}{M_P} \partial_\mu a \partial^\mu a\,,
\ee
which encodes the coupling of the volume modulus to its axionic partner. Integrating by parts and using the equation of motion 
$\Box \hat \Phi = - m_\Phi^2 \hat \Phi$ we obtain the coupling:
\be
\mathcal{L}_{\Phi a a} = \frac{1}{\sqrt{6}}\frac{m_\Phi^2}{M_P} \hat\Phi a a\,,
\label{lphiaa}
\ee
which yields the decay rate~\eqref{ddr}.

\subsubsection{Decays into visible sector fields}

The dominant volume modulus decays into visible sector \textit{dof} are to Higgses via the GM coupling $Z$. Additional leading order decay channels can be to SUSY scalars and Higgsinos depending respectively on $m_0$ and $\hat\mu$. On the other hand, as explained in \cite{DR1,DR2}, $\tau_b$ decays into visible gauge bosons are loop suppressed, i.e. $c_{\Phi\to A A}\sim \alpha_\SM^2 \ll 1$, whereas decays into matter fermions and gauginos are chirality suppressed, i.e. $c_{\Phi\to f f}\sim \left(m_f/m_\Phi\right)^2 \ll 1$. The main goal of this section is to compute the cubic interaction Lagrangian which gives rise to the decay of the volume modulus into Higgses, Higgsinos, squarks and sleptons.

\subsubsection*{Decay into scalar fields}

Let us first focus on the volume modulus decays into visible scalar fields which are induced by the $\tau_b$-dependence of both kinetic and mass terms in the total effective Lagrangian $\mc{L} = \mc{L}_{\rm kin} - V$. $\mc{L}_{\rm kin}$ is determined by the leading order K\"ahler potential:
\be
K \simeq - 3 \ln\left(\frac{T_b+\overline{T}_b}{2}\right) + \frac{2}{T_b + \overline{T}_b} \left[f_\alpha(U,S) \overline{C}^{\overline{\alpha}} C^\alpha 
+ \left(Z H_u H_d + \rm h.c.\right)\right]\,,
\ee
where we included only the leading term of the K\"ahler matter metric $\tilde{K}_\alpha$ in~\eqref{kmatter}. 
Writing each complex scalar field as $C^\alpha = \frac{\rm{Re} C^\alpha + i \rm{Im} C^\alpha}{\sqrt{2}}$, 
the canonically normalised real scalar fields look like:
\bea
h_1 &=& \lambda_u {\rm Re} H_u^+ \qquad h_2 = \lambda_d {\rm Re} H_d^-
\qquad h_3 = \lambda_d {\rm Re} H_d^0 \qquad h_4 = \lambda_u {\rm Re} H_u^0  \nn \\
h_5 &=& \lambda_d {\rm Im} H_d^0  \qquad h_6 = \lambda_u {\rm Im} H_u^0 
\qquad h_7 = \lambda_u {\rm Im} H_u^+  \qquad h_8 = \lambda_d {\rm Im} H_d^-  \nn \\
\sigma_\alpha &=& \lambda_\alpha {\rm Re} C_\alpha \qquad \chi_\alpha = \lambda_\alpha {\rm Im} C_\alpha
\qquad\text{where}\qquad \lambda_i\equiv \sqrt{\frac{f_i(U,S)}{\langle \tau_b\rangle}}\,. 
\label{fields}
\eea
Keeping only terms which are at most cubic in the fields and neglecting axion-scalar-scalar interactions, we can schematically 
write the kinetic Lagrangian as $\mc{L}_{\rm kin} = \mc{L}_{\rm{kin, quad}} + \mc{L}_{\rm{kin, cubic}}$ where:
\be
\mc{L}_{\rm{kin, quad}} = \frac 12 \partial_\mu \hat\Phi \partial^\mu \hat\Phi + \frac 12 \partial_\mu a \partial^\mu a 
+ \frac 12 \partial_\mu h_i \partial^\mu h^i + \frac{1}{2} \partial_\mu \sigma_\alpha \partial^\mu \sigma^\alpha 
+ \frac{1}{2} \partial_\mu \chi_\alpha \partial^\mu \chi^\alpha\,, \nn
\ee
while the cubic part can be further decomposed as $\mc{L}_{\rm{kin, cubic}} = \mc{L}_{\Phi a a} + \mc{L}_{\Phi h h} + \mc{L}_{\Phi C C}$, 
with $\mc{L}_{\Phi aa}$ given in~\eqref{lphiaa} and:
\be
\mc{L}_{\Phi h h} =  - \frac{1}{M_P\sqrt{6}} \left[\partial_\mu \hat\Phi h_i\partial^\mu h^i + \hat\Phi \partial_\mu h_i \partial^\mu h^i
+Z\partial_\mu \hat\Phi \sum_{i=1}^4 (-1)^{i+1}\left(h_{2i}\partial^\mu h_{2i-1} + h_{2i-1}\partial^\mu h_{2i} \right)\right], \nn
\ee
and:
\be
\mc{L}_{\Phi C C} = - \frac{1}{M_P\sqrt{6}} \left(\sigma_\alpha \partial_\mu \sigma^\alpha \partial^\mu \hat\Phi + \chi_\alpha \partial_\mu \chi^\alpha \partial^\mu \hat\Phi + \hat\Phi \partial_\mu \sigma_\alpha \partial^\mu \sigma^\alpha + \hat\Phi \partial_\mu \chi_\alpha \partial^\mu \chi^\alpha\right)\,. \nn
\ee
In addition to the LVS part, the scalar potential contains also the following terms:
\be
\label{sp1}
V \supset \frac{1}{2}  m^2_0 \left( \sigma_\alpha\sigma^\alpha + \chi_\alpha\chi^\alpha \right) 
+ \frac 12 \left(\hat{\mu}^2+m_0^2\right) h_i h^i + B\hat{\mu} \sum_{i=1}^4 (-1)^{i+1}h_{2i-1}h_{2i}\,.
\ee
Since the soft-terms depend on the volume modulus, we can expand them as:
\bea
\label{stexp}
\hat{\mu}^2&\propto& \tau_b^{-\alpha}\quad\Rightarrow\quad \hat{\mu}^2(\hat\Phi) = \hat{\mu}^2 \left(1-\alpha \sqrt{\frac 23}\frac{\hat\Phi}{M_P}\right)\,, \nn \\
m_0^2&\propto& \tau_b^{-\beta}\quad\Rightarrow\quad m_0^2(\hat\Phi) = m_0^2 \left(1-\beta \sqrt{\frac 23}\frac{\hat\Phi}{M_P}\right)\,, \nn \\
B\hat{\mu} &\propto& \tau_b^{-\gamma}\quad\Rightarrow\quad B\hat{\mu} (\hat\Phi) = B\hat{\mu} \left(1-\gamma \sqrt{\frac 23}\frac{\hat\Phi}{M_P}\right)\,,
\eea
where $\alpha$, $\beta$ and $\gamma$ depend on the specific scenario. This expansion leads to new cubic interactions coming from the terms of the scalar potential in~\eqref{sp1}:
\be
V\supset - \frac{1}{M_P\sqrt{6}} \left[\gamma m_0^2 \, \hat\Phi \left(\sigma_\alpha\sigma^\alpha  + \chi_\alpha\chi^\alpha\right) 
+\left(\alpha \hat{\mu}^2 + \beta m_0^2 \right) \, \hat \Phi h_i h^i +2 \gamma B\hat{\mu} \, \hat \Phi \sum_{i=1}^4 (-1)^{i+1}h_{2i-1}h_{2i}\right]. \nn
\ee
Including the relevant cubic interactions coming from the kinetic Lagrangian and integrating by parts, we obtain a total cubic Lagrangian of the form:
\bea
\mc{L}_{\rm cubic} &=& \frac{1}{M_P\sqrt{6}} \left[ \hat \Phi h_i \Box h^i + \hat \Phi \left(\sigma_\alpha \Box \sigma^\alpha + \chi_\alpha \Box \chi^\alpha \right) + \left(\alpha \hat{\mu}^2 + \beta m_0^2 \right) \, \hat \Phi h_i h^i\right. + \nn \\
&+& \left. \gamma m_0^2 \, \hat \Phi \left(\sigma_\alpha\sigma^\alpha + \chi_\alpha\chi^\alpha\right)+\left(Z \Box \hat \Phi + 2\gamma B\hat{\mu} \, \hat\Phi\right) \sum_{i=1}^4 (-1)^{i+1}h_{2i-1}h_{2i}\right]. \nn
\eea
The leading order expressions of the equations of motion are:
\bea
\Box \sigma_\alpha &=& - m_0^2 \sigma_\alpha \qquad 
\Box h_{2i-1} =  - \left(\hat{\mu}^2 + m_0^2 \right) h_{2i-1} + (-1)^i B\hat{\mu} \, h_{2i} \quad i=1,\dots,4 \nn \\
\Box \chi_\alpha &=& - m_0^2 \chi_\alpha \qquad
\Box h_{2j} =  - \left(\hat{\mu}^2 + m_0^2 \right) h_{2j} + (-1)^j B\hat{\mu} \, h_{2j-1} \quad j=1,\dots,4 \,, \nn
\eea
which have to be supplemented with:
\be
\Box \hat \Phi = - m_{\Phi}^2 \hat \Phi \qquad \qquad \Box a = - m_a^2 a \simeq 0\,.
\ee
Plugging these equations of motion into $\mc{L}_{\rm cubic}$, the final result becomes:
\bea
\mc{L}_{\rm cubic} &=& - \frac{1}{M_P\sqrt{6}} \left[\left(\hat{\mu}^2 \left(1 - \alpha\right) + m_0^2 \left(1 - \beta \right)\right) \, \hat\Phi h_i h^i 
+ (1 - \gamma) m_0^2 \, \hat \Phi \left(\sigma^\alpha\sigma_\alpha + \chi^\alpha\chi_\alpha\right)\right. \nn \\
&+& \left. \left(2 B\hat{\mu} \left(1 - \gamma \right) + Z\, m_\Phi^2 \right)\, \hat\Phi \sum_{i=1}^4 (-1)^{i+1}h_{2i-1}h_{2i}\right],
\label{finallagr}
\eea
from which it is easy to find the corresponding decay rates using the fact that: 
\bea
\mc{L}_1 &=& \lambda \frac{m^2}{M_P}\Phi \phi \phi \quad\Rightarrow \quad
\Gamma_{\Phi \rightarrow \phi \phi} = \lambda_1 \Gamma_0 \,,\quad \lambda_1 = 6\lambda^2\left(\frac{m}{m_{\Phi}}\right)^4 \sqrt{1 - 4 \left(\frac{m_\phi}{m_{\Phi}}\right)^2}\,, \label{dw1} \\
\mc{L}_2 &=& \lambda \frac{m^2}{M_P}\Phi \phi_1 \phi_2\,\,\Rightarrow \,\,
\Gamma_{\Phi \rightarrow \phi_1 \phi_2} = \lambda_2 \Gamma_0\,, \quad \lambda_2 = \frac{\lambda_1}{2}\,\,\text{for}\,\, m_{\phi_1} = m_{\phi_2} = m_\phi\,.
\label{dw2}
\eea

\subsubsection*{Decay into Higgsinos}

The decay of the volume modulus into Higgsinos is determined by expanding the Higgsino kinetic and mass terms around the VEV of $\tau_b$ and then working with canonically normalised fields. The relevant terms in the low-energy Lagrangian are:
\be
\mc{L} \supset i \tilde{H}^{\dagger}_i \overline{\sigma}^\mu \partial_\mu \tilde{H}^{i} \left(1-\sqrt{\frac 23}\frac{\hat\Phi}{M_P}\right)- \frac{\hat\mu}{2} \left(\tilde{H}_u^+ \tilde{H}_d^- - \tilde{H}_u^0 \tilde{H}_d^0\right)\left(1-\frac{\alpha}{\sqrt{6}} \frac{\hat\Phi}{M_P}\right) +\text{h.c.} \,.
\ee
After imposing the equations of motion, we get the following cubic interaction Lagrangian:
\be
\mc{L}_{\rm cubic} \supset \, \frac{\alpha}{2 \sqrt{6}} \, \frac{\hat\mu}{M_P} \,\hat\Phi \left(\tilde{H}_u^+ \tilde{H}_d^- - \tilde{H}_u^0 \tilde{H}_d^0\right)+\text{h.c.}\,.
\label{higgsinocubic}
\ee
The corresponding decay rates take the form:
\be
\Gamma_{\Phi \rightarrow \tilde{H}_u^+ \tilde{H}_d^-} = \Gamma_{\Phi \rightarrow \tilde{H}_u^0 \tilde{H}_d^0} = \frac{\alpha^2}{4} \left(\frac{\hat\mu}{m_\Phi}\right)^2 \left(1 - 4 \left(\frac{\hat\mu}{m_\Phi}\right)^2\right)^{3/2} \Gamma_0 \,.
\label{higgsinosdr}
\ee

\subsection{Dark radiation predictions}

It is clear from~\eqref{finallagr} and~\eqref{higgsinosdr} that the volume modulus branching ratio into visible sector \textit{dof} depends on the size of the soft-terms. Hence the final prediction for dark radiation production has to be studied separately for each different visible sector construction. 

\subsubsection{MSSM-like case}

Firstly we consider MSSM-like models arising from the ultra-local dS$_2$ case where all soft-terms are suppressed relative to the volume modulus mass:
\be
m_0^2 \simeq M_{1/2}^2 \simeq B\hat{\mu} \simeq \hat{\mu}^2 \sim \frac{M_P^2}{\vo^4} \ll m_{\Phi}^2 \sim \frac{M_P^2}{\V^3}\,.
\ee
Let us briefly review the results for dark radiation production which for this case have already been studied in \cite{DR1, DR2}. 

Given that all soft-terms are volume-suppressed with respect to $m_\Phi$, only the last term in~\eqref{finallagr} gives a non-negligible contribution to the volume modulus branching ratio into visible sector fields. Thus the leading $\Phi$ decay channel is to MSSM Higgses via the GM coupling. Using~\eqref{dw2}, we find:  
\be
\label{phihh1}
\Gamma_{\Phi \rightarrow hh} = c_{\rm vis}\Gamma_0\qquad\text{with}\qquad c_{\rm vis}= 4\times\frac{Z^2}{2} \sqrt{1 - 4\,\frac{(\hat\mu^2+m_0^2)}{m_\Phi^2}} \simeq 2 Z^2\,.
\ee
Plugging this value of $c_{\rm vis}$ together with $c_{\rm hid}=1$ (see~\eqref{ddr}) into the general expression~\eqref{dn} for extra dark radiation, we obtain the window:
\be
\frac{1.53}{Z^2}\lesssim \Delta N_{\rm eff}\lesssim  \frac{1.60}{Z^2}\,,
\label{MSSMpred}
\ee
for $0.2 \, {\rm GeV} \lesssim T_{\rm rh} \lesssim 10 \, {\rm GeV}$. Clearly this gives values of $\Delta N_{\rm eff}$ larger than unity for $Z = 1$. Using the bound~\eqref{cbound}, we see that we need $c_{\rm vis}\gtrsim 3$ in order to be consistent with present observational data, implying $Z\gtrsim \sqrt{3/2}\simeq 1.22$.

\subsubsection{Split SUSY-like case}
\label{splitsusycase}

Let us now analyse dark radiation predictions for split SUSY-like scenarios arising in the dS$_1$ (both local and ultra-local) and local dS$_2$ cases.
In these scenarios the hierarchy among soft-terms is (considering $\mu$ and $B\mu$-terms generated by $K$):
\be
\label{hierarchyss}
M_{1/2}^2 \simeq \hat{\mu}^2 \sim \frac{M_P^2}{\vo^4} \ll m_0^2 \simeq B\hat\mu \simeq m_{\Phi}^2 \sim \frac{M_P^2}{\vo^3}\,.
\ee
The main difference with the MSSM-like case is that now $B\hat\mu$ and $m_0^2$ scale as $m_\Phi^2$. In order to understand if volume modulus decays into SUSY scalars are kinematically allowed, i.e. $R~\equiv m^2_0/m_\Phi^2 \leq 1/4$, we need therefore to compute the exact value of $m_\Phi$ and compare it with the results derived in Sec.~\ref{Sec:m0}. It turns out that $m_\Phi$ depends on the dS mechanism as follows:
\be
{\rm dS}_1: \,\, m_\Phi^2 = \frac{9}{8 a_s\tau_s} \frac{m_{3/2}^2\tau_s^{3/2}}{\vo} \qquad\qquad 
{\rm dS}_2: \,\, m_\Phi^2 = \frac{27}{4 a_s\tau_s} \frac{m_{3/2}^2\tau_s^{3/2}}{\vo}\,.
\label{vmm}
\ee
\newpage
Let us analyse each case separately:
\bi
\item \emph{Local and ultra-local dS$_1$ cases}: Even if the F-term contribution to scalar masses~\eqref{lsm} for the local case can be made small by appropriately tuning the coefficient $c_s$, the D-term contribution to $m_0^2$ given by~\eqref{ul1} cannot be tuned to small values once the requirement of a dS vacuum is imposed. Hence for both local and ultra-local cases, $m_0^2$ cannot be made smaller than~\eqref{ul1} giving:
\be
m_0^2\geq 2 m_\Phi^2 \qquad\Rightarrow\qquad R \geq 2\,,
\ee
which is clearly in contradiction with the condition $R\leq 1/4$ that has to be satisfied to open up the decay channel of $\Phi$ into SUSY scalars. Therefore 
the decay of $\Phi$ into squarks and sleptons is kinematically forbidden. 

\textit{Decay to Higgses}

Similarly to the MSSM-like case, $\Phi$ can still decay to Higgs bosons via the GM term in (\ref{finallagr}). Given that $m_\Phi< m_0$, when $\Phi$ decays at energies of order $m_\Phi$, EWSB has already taken place at the scale $m_0$.\footnote{As we shall show later on, RG flow effects do not modify these considerations qualitatively.} The gauge eigenstates $h_i$ $i=1,...,8$ given in~\eqref{fields} then get rotated into $8$ mass eigenstates. $4$ Higgs \textit{dof} which we denote by $A^0, H^0, H^\pm$ remain heavy and acquire a mass of order $m_{H_d}^2 \simeq m_0^2$, and so the decay of $\Phi$ into these fields is kinematically forbidden. The remaining $4$ \textit{dof} are the $3$ would-be Goldstone bosons $G^0$ and $G^{\pm}$ which become the longitudinal components of $Z^0$ and $W^{\pm}$, and the ordinary SM Higgs field $h^0$. The $\Phi$ decay rate into light Higgs \textit{dof} can be obtained from the last term in~\eqref{finallagr} by writing the gauge eigenstates in terms of the mass eigenstates as \cite{Martin:1997ns}:
\bea
h_1 = {\rm Re} G^+ \sin\beta + {\rm Re} H^+ \cos\beta\,,&\qquad& h_2 = - {\rm Re} G^+ \cos\beta + {\rm Re} H^+ \sin\beta\,, \nn \\
h_3 = \sqrt{2} v_d + h^0 \sin\beta + H^0 \cos \beta \,, &\qquad& h_4 = \sqrt{2} v_u + h^0 \cos\beta - H^0 \sin\beta \,, \nn \\
h_5 = - G^0 \cos\beta\, + A^0 \sin \beta\,,&\qquad& h_6 = G^0 \sin\beta\, + A^0 \cos \beta\,, \nn \\
h_7 = {\rm Im} G^+ \sin\beta + {\rm Im} H^+ \cos \beta\,,&\qquad& h_8 = {\rm Im} G^+ \cos\beta - {\rm Im} H^+ \sin\beta\,,
\label{gauge-mass}
\eea
where $v_u \equiv \langle H^0_u \rangle$, $v_d \equiv \langle H^0_d \rangle$ and $\tan \beta~\equiv v_u/v_d$. Since in split SUSY-like models $\tan \beta \sim \mc{O}(1)$ in order to reproduce the correct Higgs mass \cite{Giudice:2011cg, Bagnaschi:2014rsa}, the interaction Lagrangian simplifies to:
\bea
\mc{L}_{\rm cubic} &\supset& \frac{Z}{2\sqrt{6}}\, \frac{m_\Phi^2}{M_P} \, 
\hat\Phi \left[\left(h^0\right)^2+\left(G^0\right)^2+\left({\rm Re}G^+\right)^2+\left({\rm Im} G^+\right)^2 - \right. \nn \\
&&\left.- \left(A^0\right)^2 - \left(H^0\right)^2 - \left({\rm Re}H^+\right)^2 - \left({\rm Im}H^+\right)^2\right]\,.
\label{ewsbint}
\eea
Neglecting interaction terms involving heavy Higgses, (\ref{ewsbint}) gives a decay rate of the form:
\be
\Gamma^{\rm (GM)}_{\Phi \rightarrow hh,GG}=d_1 \Gamma_0\qquad\text{with}\qquad d_1 \simeq 4\times\frac{Z^2}{4}\simeq Z^2\,.
\ee
\newpage
\textit{Decay to Higgsinos}

$\Phi$ can also decay to Higgsinos via the interaction Lagrangian (\ref{higgsinocubic}). We need only to check if this decay is kinematically allowed. In split SUSY-like models EWSB takes place at the scale $m_0$ where the following relations hold:
\be
\hat\mu^2 = \frac{m_{H_d}^2 - m_{H_u}^2 \tan^2 \beta}{\tan^2 \beta - 1} - \frac{m_Z^2}{2}\,, \qquad \quad \sin\left(2 \beta\right) = \frac{2 |B \hat\mu|}{m_{H_d}^2 + m_{H_u}^2 + 2 \hat\mu^2}\,.
\label{ewsb}
\ee
In the case of universal boundary conditions for the Higgs masses, i.e. $m_{H_u}=m_{H_d}=m_0$ at the GUT scale, the $\hat\mu$-term has necessarily to be of order the scalar masses $m_0$ since for $\tan \beta \sim \mathcal{O}\left(1\right)$ the first EWSB condition in (\ref{ewsb}) simplifies to:
\be
\hat\mu^2 \simeq \frac{m_{H_d}^2}{\tan^2 \beta - 1} \simeq m_0^2 \,,
\label{ewsb2}
\ee
given that $m_{H_u}^2$ runs down to values smaller than $m_{H_d}^2$ due to RG flow effects. On the other hand, $\hat\mu$ could be much smaller than $m_0$ for non-universal boundary conditions, i.e. if $m_{H_u} \neq m_{H_d}$ at the GUT scale. In fact, in split SUSY models $m_{H_u}^2$ is positive around the scale $m_0$, and so the first EWSB condition in (\ref{ewsb}) for $\hat\mu \ll m_0$ becomes:
\be
m_{H_d}^2 \simeq m_{H_u}^2 \tan^2 \beta \,.
\ee
This condition can be satisfied at the scale $m_0$ for a proper choice of boundary conditions at the GUT scale with $m_{H_u}>m_{H_d}$. 
Let us point out that, if $\hat\mu$ is determined by K\"ahler potential contributions as in (\ref{bmu}), $\hat\mu$ is suppressed with respect to $m_0$ but, if $\hat\mu$ is generated by non-perturbative effects in $W$, $\hat\mu$ can be of order $m_0$. In this case, the parametrisation of the $\hat\mu$-term (\ref{stexp}) reproduces the correct $\tau_b$-dependence of the non-perturbatively induced $\hat\mu$-term (\ref{mubmu}) if $\alpha = 3 n + 1$.\footnote{If the instanton number is $n = 1$, the $\hat\mu$-term can easily be of order $m_\Phi$ since $c_{\mu,W}$ is a flux-dependent tunable coefficient. For example, setting $\xi = a_s = 1$ and $g_s = 0.1$, the requirement $\hat{\mu} \simeq m_\Phi$ implies that $c_{\mu,W} \simeq W_0/(4\,\vo^{1/6})$. For $W_0 \simeq 10$ and $\vo \simeq 10^7$ we get $c_{\mu,W} \simeq 0.2$.} If we parameterise the ratio between $\hat\mu$ and $m_{\Phi}$ as $\tilde{c} \equiv \hat\mu/m_\Phi$, the decay of $\Phi$ into Higgsinos is kinematically allowed only for $\tilde{c} \leq 1/2$. Using (\ref{higgsinosdr}), this decay rate takes the form (where we set $n = 1\,\Leftrightarrow\,\alpha = 4$):
\be
\Gamma_{\Phi \rightarrow \tilde{H} \tilde{H}} = d_2 \Gamma_0 \qquad\text{with}\qquad d_2 \simeq 8 \tilde{c}^2 \left(1 - 4 \tilde{c}^2\right)^{3/2}\,.
\label{higgsinogamma}
\ee

\textit{Dark radiation prediction}

Plugging $c_{\rm vis} = d_1 + d_2$ into~\eqref{dn} with $c_{\rm hid}=1$ we get the following general result:
\be
\frac{3.07}{Z^2 + d_2}\lesssim \Delta N_{\rm eff}\lesssim  \frac{3.20}{Z^2 + d_2}\,.
\label{dS1pred}
\ee
Considering $\tilde{c} = 1/\sqrt{10}$ which maximises $d_2 \simeq 0.37$ we find that for $0.2 \, {\rm GeV} \lesssim T_{\rm rh} \lesssim 10 \, {\rm GeV}$ this prediction yields values of $\Delta N_{\rm eff}$ larger than unity for $Z = 1$. Consistency with present observational data, i.e. $\Delta N_{\rm eff}\lesssim 1$, requires $Z\gtrsim 1.68$.

\item \emph{Local dS$_2$ case}: The situation seems more promising in this case since in the local limit the D-term contribution to $m_0^2$ is volume-suppressed with respect to $m_\Phi^2$ since it scales as $\left. m_0^2\right|_D \sim \mc{O}\left(\vo^{-4}\right)$ \cite{SoftTermsSeqLVS}. In this case it is therefore possible to tune the coefficient $c_s$ to obtain $R\leq 1/4$. By comparing the second term in~\eqref{vmm} with~\eqref{lsm}, this implies that $c_s$ has to be tuned so that $\left(c_s - \frac 13\right) \leq \frac{9}{10 \, a_s \tau_s}$, where $a_s \tau_s \sim 80$ in order to get TeV-scale gaugino masses \cite{SoftTermsSeqLVS}. However the condition $m_0^2 > 0$ to avoid tachyonic masses translates into $\left(c_s - \frac 13\right)>0$, giving rise to a very small window: 
\be
0 < \left( c_s- \frac 13 \right) \leq \frac{9}{10 \, a_s \tau_s} \simeq 0.01\,.
\ee
Given that $c_s$ should be extremely fine-tuned, it seems very unlikely to open this decay channel. However the total K\"ahler potential, 
on top of pure $\alpha'$ corrections, can also receive perturbative string loop corrections of the form \cite{Berg:2007wt}:
\be
K_{\rm loop} = \frac{g_s C_{\rm loop}}{\vo^{2/3}}\left(1+ k_{\rm loop}\sqrt{\frac{\tau_s}{\tau_b}}\right)\,,
\label{Kloop}
\ee
where $C_{\rm loop}$ and $k_{\rm loop}$ are two $\mc{O}(1)$ coefficients which depend on the complex structure moduli. Due to the \textit{extended no-scale structure} \cite{extendednoscale}, $g_s$ effects do not modify the leading order scalar potential, and so the mass of the volume modulus is still given by~\eqref{vmm}. However, in order to reproduce a correct ultra-local limit~\eqref{ul}, we need to change the parametrisation of the K\"ahler matter metric from~\eqref{Ktilde} to:
\be
\tilde{K} = \frac{1}{\vo^{2/3}} \left(1 - c_s \frac{\hat{\xi}}{\vo} - c_{\rm loop} \frac{g_s C_{\rm loop}}{\vo^{2/3}} \right),
\label{newKtilde}
\ee
where we introduced a new coefficient $c_{\rm loop}$ and we neglected $k_{\rm loop}$-dependent corrections in~\eqref{Kloop} since they are subdominant in the large volume limit $\tau_s\ll\tau_b$. The new ultra-local limit is now given by $c_s = c_{\rm loop} = 1/3$. 

\begin{figure}[!ht]
\begin{center}
\includegraphics[width=0.5\textwidth, angle=0]{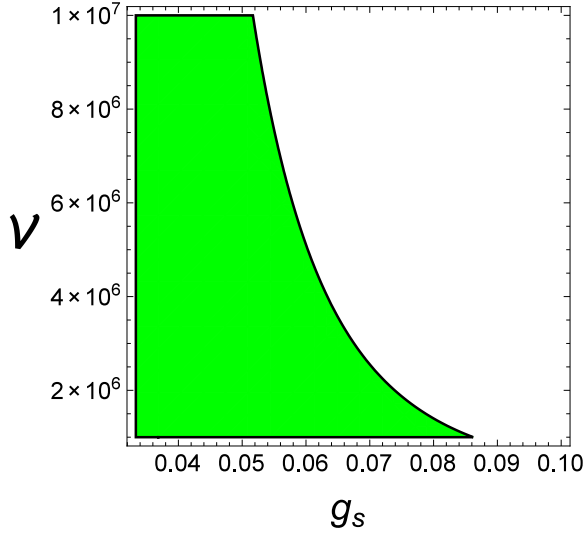}
\caption{The green region in the ($g_s$, $\vo$) parameter space gives $R \leq 1/4$, and so the decay channel of $\Phi$ into SUSY scalars is kinematically allowed.} 
\label{psf}
\end{center}
\end{figure}

These new $c_{\rm loop}$-dependent corrections in~\eqref{newKtilde} affect the final result for scalar masses and can therefore open up the $\Phi$ decay channel to SUSY scalars. In fact, the result~\eqref{lsm} for scalar masses in the local case gets modified to: 
\be
\left.m_0^2\right|_F = \frac{15}{2}\, \frac{m_{3/2}^2 \tau_s^{3/2}}{\vo}
\left[\left(c_s-\frac{1}{3}\right) - \frac{8 g_s C_{\rm loop}}{15} \left(c_{\rm loop} - \frac 13\right)
\frac{\vo^{1/3}}{\hat\xi}\right] \,.
\label{newlsm}
\ee
The two terms in square brackets are of the same order for $g_s\simeq 0.1$ and $\vo \sim 10^7$ which is needed to get TeV-scale gauginos, and so they can compete to get $R \leq \frac 14$. As an illustrative example, if we choose $c_s = 1/3$ and natural values of the other parameters: $C_{\rm loop} = a_s = 1$, $\xi = 2$ and $c_{\rm loop} = 0$ (non-tachyonic scalars require $c_{\rm loop} < 1/3$ for $c_s = 1/3$), the ratio between squared masses becomes:
\be
R = \frac{8}{81}\, g_s^{3/2}\,\vo^{1/3}\,.
\ee
As can be seen from Fig.~\ref{psf}, there is now a wide region of the parameter space where the $\Phi$ decay to SUSY scalars is allowed.

We finally point out that $g_s$ corrections to the K\"ahler matter metric affect the result for scalar masses only in the local case since in the ultra-local limit $m_0$ is generated by effects (D-terms for dS$_1$ and F-terms of $S$ and $U$-moduli for dS$_2$) which are sensitive only to the leading order expression of $\tilde{K}_\alpha$.
\ei
Let us now analyse the final prediction for dark radiation production for split SUSY-like models where the decay channel of $\Phi$ into SUSY scalars is kinematically allowed.

\subsubsection*{Dark radiation results}
\label{drresults}

We start by parameterising the scalar mass $m_0$ in terms of the volume mode mass $m_\Phi$ as $m_0 = c \,m_\Phi$ and the $\hat\mu$-term as $\mu = \tilde{c}\,m_\Phi$ so that the corresponding kinematic constraints for $\Phi$ decays into SUSY scalars and Higgsinos become $c \leq \frac 12$ and $\tilde{c}\leq \frac 12$.
Parameterising also the $B\hat\mu$-term as in (\ref{bmu}) and using the fact that for split SUSY-like models we have in (\ref{stexp}) $\beta = \gamma = 9/2$ and $\alpha = 4$,\footnote{We focus on the case where the $\hat\mu$-term is generated by non-perturbative effects in $W$ since when $\hat\mu$ is generated by $K$, it turns out to be very suppressed with respect to $m_0$, i.e. $\tilde{c}\ll 1$, and so it gives rise to a negligible contribution to the branching ratio of $\Phi$.} the leading order cubic Lagrangian is given by the sum of (\ref{finallagr}) and (\ref{higgsinocubic}):
\bea
\mc{L}_{\rm cubic} &\simeq& \frac{7 c^2}{2\sqrt{6}}\frac{m_\Phi^2}{M_P} \hat\Phi\left[  
\sigma^\alpha\sigma_\alpha + \chi^\alpha\chi_\alpha+\left(1+\frac{6\tilde{c}^2}{7 c^2}\right)h_i h^i+ 2Z  \left(c_{B,K}  - \frac{1}{7 c^2} \right) \sum_{i=1}^4 (-1)^{i+1}h_{2i-1}h_{2i}\right] \nn \\
&+& \tilde{c}\sqrt{\frac 23} \, \frac{m_\Phi}{M_p}\hat\Phi \left(\tilde{H}_u^+\tilde{H}_d^- -\tilde{H}_u^0\tilde{H}_d^0\right)+\text{h.c.}.
\label{intlagr}
\eea
Contrary to the MSSM-like case, now the decay of the volume modulus into squarks and sleptons through mass terms is kinematically allowed and also the decay rate into Higgses is enhanced due to mass terms and $B\hat\mu$ couplings. Using~\eqref{dw1}, the total decay rate into squarks and sleptons reads:
\be
\Gamma_{\Phi \rightarrow \sigma\sigma,\chi\chi} = c_0 \Gamma_0\qquad\text{with}\qquad c_0 = N\times \frac{49 \, c^4}{4} \sqrt{1 - 4 c^2}\,, 
\ee
where $N=90$ is the number of real scalar \textit{dof} of the MSSM,\footnote{$12$ \textit{dof} for each left handed squark doublet ($3$ families), $6$ \textit{dof} for each right handed squark ($6$ squarks), $4$ \textit{dof} for each left handed slepton doublet ($3$ families), $2$ \textit{dof} for each right handed slepton ($3$ families).} except for the Higgses.

\begin{figure}[!ht]
\begin{center}
\includegraphics[width=0.42\textwidth, angle=0]{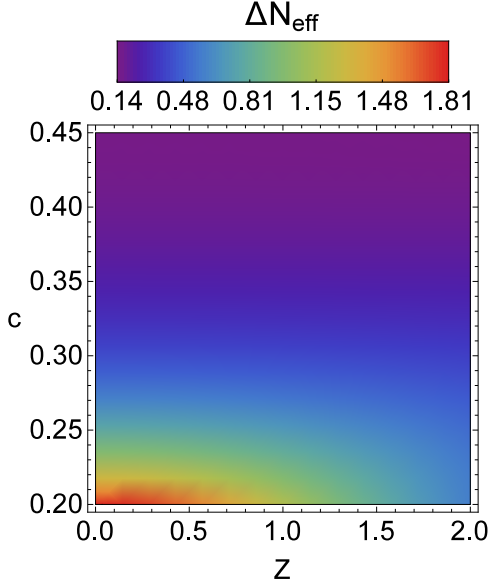}
\caption{Dark radiation production in split SUSY-like models for $T_{\rm rh} \gtrsim 0.7 \, {\rm GeV}$ and $\tilde{c}=0$.} 
\label{ris1}
\end{center}
\end{figure}

On the other hand, the decay rate into Higgs bosons receives contributions from both mass and GM terms. Using~\eqref{dw1} we obtain:
\be
\Gamma^{\rm (mass)}_{\Phi \rightarrow hh} = c_1 \Gamma_0\qquad\text{with}\qquad c_1 = 8 \times \frac{49 \, c^4}{4} \left(1+\frac{2\tilde{c}^2}{7 c^2}\right)^2\sqrt{1 - 4 (c^2+\tilde{c}^2)}\,, 
\ee
where $8$ is the number of MSSM real Higgs \textit{dof} while using (\ref{bmu}) and (\ref{dw2}) we get:
\be
\Gamma^{\rm (GM)}_{\Phi \rightarrow hh} = c_2 \Gamma_0\qquad\text{with}\qquad c_2 = 4\times\frac{Z^2}{2} \left(7 c_{B,K} c^2  - 1 \right)^2 \sqrt{1 - 4\,c^2}\,.
\ee
The decay rate into Higgsinos is given again by~\eqref{higgsinogamma} and thus the total $\Phi$ decay rate into visible sector fields becomes:
\be
\Gamma_{\rm vis} = \Gamma_{\Phi \rightarrow \sigma\sigma,\chi\chi}+\Gamma^{\rm (mass)}_{\Phi \rightarrow hh}+\Gamma^{\rm (GM)}_{\Phi \rightarrow hh}+\Gamma_{\Phi \rightarrow \tilde{H}\tilde{H}}= c_{\rm vis} \Gamma_0 \,,
\ee
\be
\text{where}\qquad c_{\rm vis} = c_0 + c_1 + c_2 + d_2 \,.
\ee
The final prediction for dark radiation production is then given by~\eqref{dn} with $c_{\rm hid}=1$, $g_*(T_{\rm dec}) = 10.75$ and $g_*(T_{\rm rh}) = 86.25$ for $T_{\rm rh} \gtrsim 0.7\,{\rm GeV}$.\footnote{The results do not change significantly for $g_*(T_{\rm rh}) = 75.75$ which is valid for $0.2\,{\rm GeV} \lesssim T_{\rm rh} \lesssim 0.7\, {\rm GeV}$.} The results are plotted in Fig.~\ref{ris1} where we have set $c_{B,K}=1$, $Z = 1$ and we are considering a conservative case in which the decay into Higgsinos is negligible, i.e. $\tilde{c} = 0$. For $c>0.2$, the vast majority of the parameter space yields $\Delta N_{\rm eff} \lesssim 1$, in perfect agreement with present experimental bounds with a minimum value $\left.\Delta N_{\rm eff}\right|_{\rm min} \simeq 0.14$ at $c \simeq 1/\sqrt{5}$.

\begin{figure}[!ht]
\begin{center}
\includegraphics[width=0.7\textwidth, angle=0]{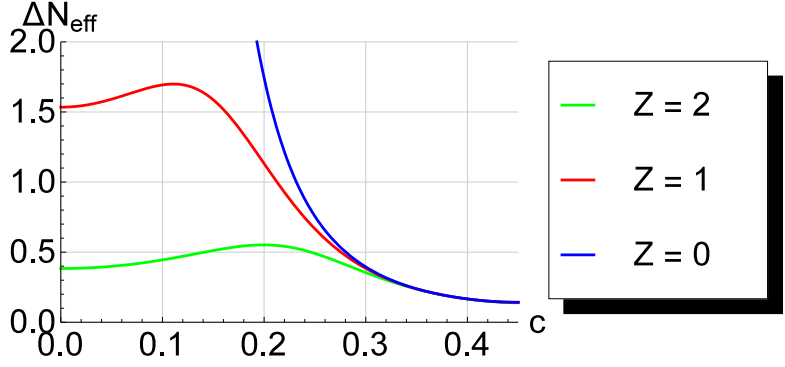}
\caption{Predictions for $\Delta N_{\rm eff}$ in split SUSY-like models with $\tilde{c}=0$ for $Z = 0$ (blue line), $Z = 1$ (red line) and $Z = 2$ (green line).} 
\label{z123}
\end{center}
\end{figure}

It is interesting to notice that, contrary to the MSSM-like case, dark radiation overprodution can now be avoided if the GM term is absent or it is very suppressed. In fact even for $Z=0$, $\Delta N_{\rm eff} \lesssim 1$ if $c\gtrsim 0.23$, as a consequence of the fact that in this region of the parameter space almost the whole suppression of $\Delta N_{\rm eff}$ is due to the decay into scalar fields. The predictions for $\Delta N_{\rm eff}$ for different values of the GM coupling $Z = 0$ (blue line), $Z = 1$ (red line) and $Z = 2$ (green line) are shown in Fig.~\ref{z123}. 

For $\tilde{c} \neq 0$ $\Delta N_{\rm eff}$ is even further suppressed than what is shown in Fig.~\ref{ris1} and~\ref{z123} but the correction is at the percent level in the interesting region where the decay into scalars dominates $\Gamma_{\rm vis}$. For example including the effect of decays into Higgsinos and setting $\tilde{c} \simeq 1/\sqrt{10}$ to maximise the decay rate into Higgsinos, the correction $\left.\delta \Delta N_{\rm eff}\right|_{\rm min}$ to $\left.\Delta N_{\rm eff}\right|_{\rm min}$ turns out to be:
\be
\frac{\left.\delta \Delta N_{\rm eff}\right|_{\rm min}}{\left.\Delta N_{\rm eff}\right|_{\rm min}} \simeq 0.03 \,.
\ee

Note that $\Delta N_{\rm eff}$ can be further suppressed by choosing $c_{B,K} > 1$, namely by enhancing the contribution due to the $B\hat\mu$-term. However 
the decay scale $m_\Phi$ is just slightly larger than the EWSB scale $m_0$ where the $4$ \textit{dof} of the two Higgs doublets get rotated into the heavy Higgs mass eigenstates $A^0, H^0, H^\pm$, the SM Higgs $h^0$ and the longitudinal components $G^0, G^\pm$ of the vector bosons $Z, W^\pm$. Hence the $B\hat\mu$-term gets reabsorbed into the mass terms for $A^0, H^0, H^\pm$ and $h^0, G^0, G^\pm$, and so varying $c_{B,K}$ does not enhance $\Gamma_{\rm vis}$ which receives its main contributions from the decay into squarks, sleptons and heavy Higgses through the mass term (assuming that is kinematically allowed) and into all Higgs \textit{dof} via the GM term. These considerations will become more clear in the next section where we will take into account corrections due to RG flow effects. A possible way-out could be the separation between the EWSB scale and the volume modulus mass, which translates into requiring $c \ll 1$. However this choice would imply $m_0^2 \ll m_\Phi^2 \sim B\hat\mu$ which would be a quite unnatural situation from a top-down perspective since $m_\Phi$ and $m_0$ have the same volume scaling.

\subsubsection*{RG flow effects}
\label{rgflows}

The results obtained above have to be corrected due to RG flow effects from the string scale to the actual scale $m_\Phi$ where the modulus decay takes place \cite{DRloop}. However these corrections do not alter qualitatively our results since $\Phi$ interacts only gravitationally and the running of squarks and sleptons in split SUSY models is almost absent \cite{Bagnaschi:2014rsa}. For sake of completeness, let us study these RG flow effects in detail. 

The soft-terms $m_0$, $\hat\mu$ and $B\hat\mu$ entering in the interaction Lagrangian~\eqref{finallagr} are just boundary conditions for the RG flow and should instead be evaluated at the scale $m_\Phi$ where the light modulus $\Phi$ decays. The RG equations for the first and second generation of squarks and sleptons are given by:
\be
m^2_\alpha = m_0^2 + \sum_{a=1}^3 c_{\alpha,a} K_a\,,
\ee
where $c_{\alpha,a}$ is the weak hypercharge squared for each SUSY scalar and the RG running contributions $K_a$ are proportional 
to gaugino masses \cite{Martin:1997ns}. Given that in split SUSY-like models gaugino masses are hierarchically lighter than scalar masses, 
the RG running of first and second generation squarks and sleptons is a negligible effect. Thus we can consider their mass at the scale $m_\Phi$ as still given by $m_0$ to a high level of accuracy. 

The situation for the third generation is slightly trickier since there are additional contributions from large Yukawa couplings. Using mSUGRA boundary conditions, the relevant RG equations become (ignoring contributions proportional to $M_{1/2}$) \cite{Martin:1997ns}:
\be
16 \pi^2 \frac{d}{dt} m_{Q_3}^2 = X_t + X_b\,,\qquad 16 \pi^2 \frac{d}{dt} m_{\ov{u}_3}^2 = 2 X_t\,,
\qquad 16 \pi^2 \frac{d}{dt} m_{\ov{d}_3}^2 = 2 X_b\,,
\label{RG1}
\ee
\be
16 \pi^2 \frac{d}{dt} m_{L_3}^2 = X_\tau\,,\qquad 16 \pi^2 \frac{d}{dt} m_{\ov{e}_3}^2 = 2 X_\tau\,,
\label{RG2}
\ee
which are coupled to those involving Higgs masses:
\be
16 \pi^2 \frac{d}{dt} m_{H_u}^2 = 3 X_t + X_b\,,\qquad 16 \pi^2 \frac{d}{dt} m_{H_d}^2 = 3 X_b + X_\tau\,.
\label{RG3}
\ee
The quantities $X_i$ look like:
\bea
\label{xi1}
& X_t = 2 |y_t|^2 \left(m_{H_u}^2 + m_{Q_3}^2 + m_{\ov{u}_3}^2\right) + 2 |a_t|^2\,, & \\
\label{xi2}
& X_b = 2 |y_b|^2 \left(m_{H_u}^2 + m_{Q_3}^2 + m_{\ov{d}_3}^2\right) + 2 |a_b|^2\,, & \\
\label{xi3}
& X_\tau = 2 |y_\tau|^2 \left(m_{H_u}^2 + m_{L_3}^2 + m_{\ov{e}_3}^2\right) + 2 |a_\tau|^2\,, &
\eea
where $y_i$ are the Yukawa couplings and $a_i$ are the only sizable entries of the $A$-term couplings. Given that for sequestered scenarios the $A$-terms scale as $M_{1/2}$ \cite{SoftTermsSeqLVS}, the contribution $2 |a_i|^2$ can be neglected with respect to the first term in each $X_i$. Moreover $X_t$, $X_b$ and $X_\tau$ are all positive, and so the RG equations~\eqref{RG1} and~\eqref{RG2} drive the scalar masses to smaller values at lower energies. This has a two-fold implication:
\bi
\item When $m_0 > m_\Phi$, RG running effects could lower $m_0$ to values smaller than $m_\Phi/2$ so that the decay channel to SUSY scalars opens up at the scale $m_\Phi$. However this never happens since RG effects are negligible.

\item When $m_0 \leq m_\Phi/2$, no one of the scalars becomes heavier than the volume modulus if $R < \frac 14$ at the boundary energy scale. On the other hand, RG running effects could still lower the scalar masses too much, suppressing the $\Phi$ decay rate to SUSY scalars. However this does not happen since RG effects are negligible.
\ei
In split SUSY-like models a correct radiative realisation of EWSB requires a low value of $\tan \beta$ \cite{Bagnaschi:2014rsa}, which implies $y_b,\,y_\tau\ll y_t$. In turn, $X_b$ and $X_\tau$ give rise to a tiny effect, and so the running of $m_{H_d}^2$, $m_{\ov{d}_3}^2$, $m_{L_3}^2$ and $m_{\ov{e}_3}^2$ turns out to be negligible. In the end, the only relevant RG equations become:
\be
16 \pi^2 \frac{d}{dt} m_{Q_3}^2 \simeq X_t\,, \qquad 16 \pi^2 \frac{d}{dt} m_{\ov{u}_3}^2 \simeq 2 X_t\,, \qquad 16 \pi^2 \frac{d}{dt} m_{H_u}^2 = 3 X_t\,.
\ee

\begin{figure}[!ht]
\begin{center}
\includegraphics[width=0.8\textwidth, angle=0]{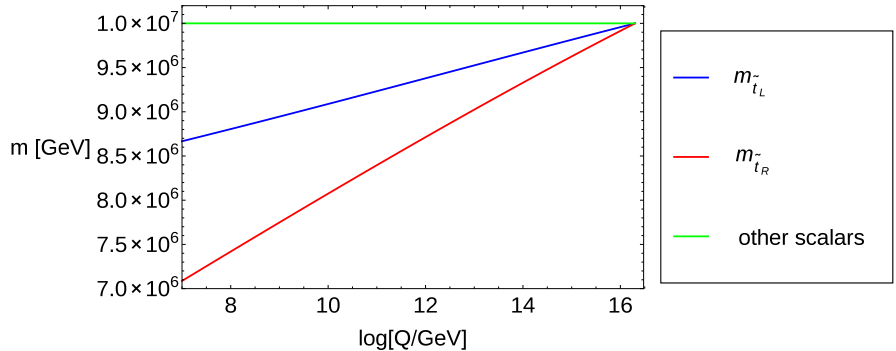}
\caption{Running of the scalar masses.\label{scr}}
\end{center}
\end{figure}

\begin{figure}[!ht]
\begin{center}
\includegraphics[width=0.8\textwidth, angle=0]{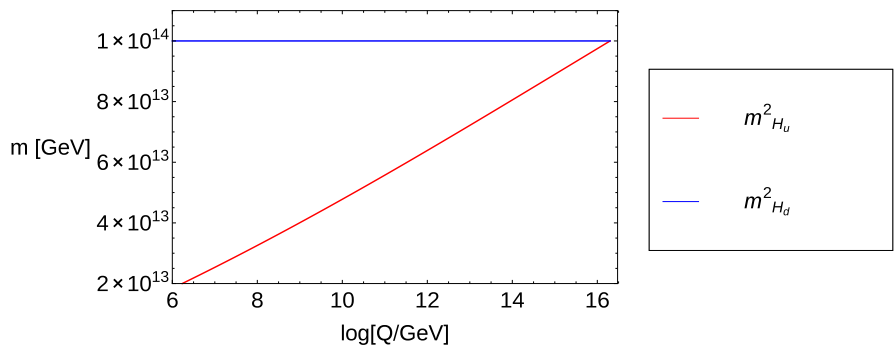}
\caption{Running of the Higgs masses.}\label{hr}
\end{center}
\end{figure}

We performed a numerical computation of the RG running, using as boundary conditions $m_0 = B\hat\mu^{1/2} = \hat\mu = 10^7 \, \rm GeV$ and $M_a = 10^3 \, \rm GeV$ at the GUT scale $M_{\rm GUT} = 2 \times 10^{16}$ GeV and $\tan \beta \simeq 1.4$. We also used that the stop left-right mixing is given by $\chi_t = A_t - \hat\mu \cot \beta \simeq - m_0/{\tan \beta}$, being $A_t \simeq M_{1/2} \ll m_0$. We used SusyHD \cite{Vega:2015fna} to run the Yukawa couplings from the top mass scale up to $m_0$ combined with SARAH \cite{Staub:2008uz} to run them from $m_0$ up to the GUT scale.\footnote{We are grateful to J. P. Vega for useful discussions about this point.} These runnings have been computed at order one loop. Using the values of the Yukawa couplings obtained at the GUT scale, we have been able to compute the running of scalars, $\hat\mu$, $B\hat\mu$ and the GM coupling $Z$ down to the scale of the decay $m_\Phi$. Fig.~\ref{scr} shows the running of scalar masses while Fig.~\ref{hr} showns the running of $m_{H_u}^2$ and $m_{H_d}^2$. The running of $\hat\mu$ and $B\hat\mu$ is almost negligible. We clarify that our purpose here is not to study EWSB in full detail but to understand which kind of behaviour we should expect for the running of soft-terms from the GUT scale to $m_\Phi$ using boundary conditions which are consistent with EWSB.

Due to RG running effects each of the scalars has a different mass at the scale $m_\Phi$, and so the exact prediction for $\Delta N_{\rm eff}$ becomes:
\be
\Delta N_{\rm eff} = \frac{43}{7} \mathcal{R} \left(\frac{g_* (T_{\rm decay})}{g_*(T_{\rm reheat})}\right)^{1/3} \,,
\ee
where:
\be
\mc{R} = \frac{\Gamma_0}{\Gamma_{\Phi \rightarrow \sigma\sigma,\chi\chi} + \Gamma_{\Phi \rightarrow hh} + \Gamma_{\Phi \rightarrow \tilde{H} \tilde{H}}}\,.
\ee
The decay rate into squarks and sleptons is given by:
\be
\label{gs}
\Gamma_{\Phi \rightarrow \sigma\sigma,\chi\chi} = \frac{49}{4} \sum_\alpha \, \kappa_\alpha \left(\frac{m_\alpha}{m_\Phi}\right)^4 \sqrt{1 - 4 \left(\frac{m_\alpha}{m_\Phi}\right)^2}\Gamma_0\,,
\ee
where the index $\alpha$ runs over all squarks and sleptons $m_\alpha = \left( m_{\tilde{Q}_{\rm L}}, m_{\tilde{u}_R}, m_{\tilde{d}_R}, m_{\tilde{L}}, m_{\tilde{e}_{\rm R}}\right)$ while $\kappa_\alpha$ is the number of \textit{dof} for each scalar.\footnote{It turns out that $\kappa_\alpha = \left(36, 18, 18, 12, 6\right)$.}

On the other hand the decay rate into Higgs \textit{dof} is given by (we focus on the case where $\hat\mu\ll m_0$ since, as we have seen in the previous section, a large $\hat\mu$-term gives rise just to a negligible correction to the final dark radiation prediction):
\be
\label{gh}
\Gamma_{\Phi \rightarrow hh} = \left[\sum_{I \in \{A^0, H^0, H^\pm\}} \left(49 \left(\frac{m_{I}}{m_{\Phi}} \right)^4 + Z^2 \right) \sqrt{1 - 4 \left(\frac{m_{I}}{m_{\Phi}}\right)^2} + Z^2\right]\Gamma_0\, .
\ee
All quantities in (\ref{gs}) and (\ref{gh}) have to be computed at the decay scale $m_\Phi$. As already explained in the previous section, $B\hat\mu$ does not explicitly contribute to (\ref{gh}) since it gets reabsorbed into the Higgs masses due to EWSB. The decay of $\Phi$ into heavy Higgses $A^0, H^0, H^\pm$ through the mass term can instead contribute to $\Gamma_{\rm Higgs}$, provided that $m_{A^0, H^0, H^\pm}/m_\Phi \leq 1/2$. The mass of the heavy Higgses in the limit $m_Z, m_{W^\pm} \ll m_{A^0}$ can be written as \cite{Martin:1997ns}:
\be
m_{H^0}^2 \simeq m_{H^\pm}^2 \simeq m_{A^0}^2 \simeq 2 |\hat\mu|^2 + m_{H_u}^2 + m_{H_d}^2 \,.
\ee
The decay rate into Higgsinos instead reads:
\be
\Gamma_{\Phi \rightarrow \tilde{H} \tilde{H}} = 8 \left(\frac{\hat\mu}{m_\Phi}\right)^2 \left(1 - 4 \left(\frac{\hat\mu}{m_\Phi}\right)^2\right)^{3/2} \Gamma_0 \,,
\label{higgsinodecayrg}
\ee
where we added the two contributions in~\eqref{higgsinosdr} and we used $\alpha = 4$. In~\eqref{higgsinodecayrg} $\hat\mu/m_\Phi \neq \tilde{c}$, since $\hat\mu$ is computed at the decay scale $m_\Phi$.

We computed $\Delta N_{\rm eff}$ for different values of $m_\Phi$, keeping the boundary conditions fixed at $m_0 = B\hat\mu^{1/2} = \hat\mu = 10^7 \, \rm GeV$ and $M_a = 10^3$ GeV. The qualitative behaviour of $\Delta N_{\rm eff}$ is the same as in the previous section where RG flow effects have been ignored. The results are shown in Tab.~\ref{tab1} which shows that the dominant contribution to $\Delta N_{\rm eff}$ is given by the decay into squarks and sleptons while the suppression coming from decay into Higgsinos is always subdominant. If $m_\Phi \simeq 2.2 \times 10^7 \, \rm GeV$, corresponding to $m_0/m_\Phi \simeq 1/\sqrt{5}$, $\Delta N_{\rm eff} \simeq 0.15$ which is only slightly larger then $\left.\Delta N_{\rm eff}\right|_{\rm min} = 0.14$ computed in the previous section without taking into account RG flow effects. This is due to the fact that the running of the SUSY scalars is negligible as can be clearly seen from Fig.~\ref{scr}.

\begin{table}[h!]
\begin{center}
\begin{tabular}{cccccc}
\hline
$m_\Phi$ & $\Gamma_{\rm scalars}/\Gamma_0$ & $\Gamma_{\rm Higgs}/\Gamma_0$ & $\Gamma_{\rm Higgsinos}/\Gamma_0$ & $\Delta N_{\rm eff}$  \\
\hline
$2.2 \times 10^7 \, \rm GeV$ & $18.53$ & $1.12 \,(*)$ & $0.08$  & $0.15$ \\
\hline
$3 \times 10^7 \, \rm GeV$ & $9.19$ & $1.12 \,(*)$ & $0.36$ & $0.29$ \\
\hline
$4 \times 10^7 \, \rm GeV$ & $3.36$ & $2.52$ & $0.33$ & $0.49$ \\
\hline
$5 \times 10^7 \, \rm GeV$ & $1.45$ & $2.45$ & $0.25$ & $0.74$ \\
\hline
\end{tabular}
\end{center}
\caption{Values of $\Delta N_{\rm eff}$ corresponding to different masses $m_\Phi$ of $\Phi$ for fixed boundary condition $m_0 = 10^7$ GeV at the GUT scale. We also indicate the relative importance of the various decay channels. In the case denoted by a $(*)$ the only non-vanishing contribution to $\Gamma_{\rm Higgs}$ is due to the decay into light Higgs \textit{dof} through the GM coupling, since the decay into heavy Higgs \textit{dof} turns out to be kinematically forbidden as a consequence of the RG flow: $2 m_{A^0, H^0, H^\pm} > m_\Phi$ at the decay scale $m_\Phi$. The decay into Higgsinos is always a subleading effect.}
\label{tab1}
\end{table}

\section{Conclusions}
\label{Concl}

Extra dark radiation is a very promising window for new physics beyond the Standard Model. Its presence is a generic feature of string models where some of the moduli are stabilised by perturbative effects since the corresponding axionic partners remain very light and can behave as extra neutrino-like species \cite{DMDRcorr}. These light hidden sector \textit{dof} are produced by the decay of the lightest modulus \cite{NTDM, NonThDMinSeqLVS} leading to $\Delta N_{\rm eff}\neq 0$ \cite{DR1, DR2, AxionProbl}. 

In this paper we performed a general analysis of axionic dark radiation production in sequestered LVS models where the visible sector is localised on D3-branes at singularities \cite{Aldazabal:2000sa, Conlon:2008wa, CYembedding}. These models yield a very interesting post-inflationary cosmological history where reheating is driven by the decay of the lightest modulus $\Phi$ with a mass of order $m_\Phi\sim 10^7$ GeV which leads to a reheating temperature of order $T_{\rm rh}\sim 1$ GeV. The gravitino mass is much larger than $m_\Phi$ ($m_{3/2}\sim 10^{10}$ GeV), so avoiding any gravitino problem \cite{gravProbl}. Low-energy SUSY can still be achieved due to sequestering effects that keep the supersymmetric partners light. Gauginos are around the TeV-scale whereas squarks and sleptons can either be as light as the gauginos or as heavy as the lightest modulus $\Phi$ depending on the moduli dependence of the matter K\"ahler metric and the mechanism responsible for achieving a dS vacuum \cite{SoftTermsSeqLVS}. 

The final prediction for dark radiation production due to the decay of the volume modulus into ultra-light bulk closed string axions depends on the details of the visible sector construction:
\ben
\item \textit{MSSM-like case}: \\
MSSM-like models arise from the ultra-local dS$_2$ case where the leading visible sector decay channel of $\Phi$ is to Higgses via the GM coupling $Z$. The simplest model with two Higgs doublets and $Z=1$ gives $1.53 \lesssim \Delta N_{\rm eff}\lesssim 1.60$ for $500\,{\rm GeV} \lesssim T_{\rm rh}\lesssim 5\,{\rm GeV}$ \cite{DR1, DR2}. Values of $\Delta N_{\rm eff}$ smaller than unity require $Z\gtrsim 1.22$ or more than two Higgs doublets.

\item \textit{Split SUSY-like case with $m_0> m_\Phi/2$}: \\
Local and ultra-local dS$_1$ cases give rise to split SUSY-like scenarios where scalar masses $m_0$ receive a contribution from D-terms which cannot be made smaller than $m_\Phi/2$. Thus the decay of $\Phi$ into squarks and sleptons is kinematically forbidden. The leading visible sector decay channel of $\Phi$ is again to Higgses via the GM coupling $Z$. However, given that EWSB takes place at the scale $m_0$ which in these cases is larger than the decay scale $m_\Phi$, the volume mode $\Phi$ can decay only to the 4 light Higgs \textit{dof}. For a shift-symmetric Higgs sector with $Z=1$, the final prediction for dark radiation production is $3.07 \lesssim \Delta N_{\rm eff}\lesssim 3.20$. Consistency with present experimental data, i.e. $\Delta N_{\rm eff}\lesssim 1$, requires $Z\gtrsim 1.68$. In most of the parameter space of split SUSY-like models a correct radiative EWSB can be achieved only if the $\hat\mu$-term is of order the scalar masses. Hence, depending on the exact value of $\hat\mu$, the decay of $\Phi$ into Higgsinos could not be mass suppressed. However it gives rise just to a negligible contribution to $\Delta N_{\rm eff}$.

\item \textit{Split SUSY-like case with $m_0\leq m_\Phi/2$}: \\
Given that in the local dS$_2$ case the D-term contribution to scalar masses is negligible, the decay of $\Phi$ into SUSY scalars can become kinematically allowed. In fact, thanks to the inclusion of string loop corrections to the K\"ahler potential \cite{Berg:2007wt, extendednoscale}, a large region of the underlying parameter space features $m_0\leq m_\Phi/2$. Hence the final prediction for $\Delta N_{\rm eff}$ gets considerably reduced with respect to the previous two cases since, in addition to decays into Higgses via the GM term, leading order contributions to the branching ratio to visible sector particles involve decays into squarks and sleptons, decays into heavy Higgses induced by mass terms and possible decays into Higgsinos depending on the exact value of the $\hat\mu$-term. Depending on the exact value of $m_0$, the simplest model with $Z=1$ gives $0.14 \lesssim \Delta N_{\rm eff} \lesssim 1.6$. Hence these models feature values of $\Delta N_{\rm eff}$ in perfect agreement with present observational bounds. Note that dark radiation overproduction can be avoided even for $Z=0$ due to the new decay channels to squarks and sleptons.
\een

We finally studied corrections to these results due to RG flow effects from the string scale $M_s \sim 10^{15}$ GeV to the volume mode mass $m_\Phi\sim 10^7$ GeV where the actual decay takes place. However these corrections do not modify our predictions since the RG running of SUSY scalar masses is a negligible effect in split SUSY-like models and radiative corrections to $m_\Phi$ are tiny since moduli are only gravitational coupled. 

\section*{Acknowledgements}

We thank George Efstathiou, Luis Aparicio and Dario Buttazzo for useful discussions. We are particularly grateful to Javier Pardo Vega for sharing with us part of the code used for the numerical computation of the RG flow.

\end{document}